\documentclass[prb,twocolumn,superscriptaddress,showpacs,floatfix]{revtex4}

\usepackage{amsfonts}
\usepackage{amssymb}
\usepackage{amsmath}
\usepackage[colorlinks=true]{hyperref}
\usepackage{graphicx}
\usepackage{tikz}
\usepackage{subfigure}

\newcommand{\vc}[1]{{\pmb{#1}}}

\begin{document}
\title{Experimental determination of Ramsey numbers}

\date{\today}

\author{Zhengbing Bian}
\affiliation{D-Wave Systems, Inc., 100-4401 Still Creek Drive,
Burnaby, British Columbia V5C 6G9, Canada}

\author{Fabian Chudak}
\affiliation{D-Wave Systems, Inc., 100-4401 Still Creek Drive,
Burnaby, British Columbia V5C 6G9, Canada}

\author{William G. Macready}
\affiliation{D-Wave Systems, Inc., 100-4401 Still Creek Drive,
Burnaby, British Columbia V5C 6G9, Canada}

\author{Lane Clark}
\affiliation{Department of Mathematics, Southern Illinois University,
Carbondale, IL 62901-4401}

\author{Frank Gaitan}
\affiliation{Laboratory for Physical Sciences, 8050 Greenmead Dr,
College Park, MD 20740}

\begin{abstract}
Ramsey theory is a highly active research area in mathematics that studies
the emergence of order in large disordered structures. Ramsey numbers mark 
the threshold at which order first appears and are extremely difficult to 
calculate due to their explosive rate of growth. Recently, an algorithm that 
can be implemented using adiabatic quantum evolution has been proposed that 
calculates the two-color Ramsey numbers $R(m,n)$. Here we present results of 
an experimental implementation of this algorithm and show that it correctly 
determines the Ramsey numbers $R(3,3)$ and $R(m,2)$ for $4\leq m\leq 8$. 
The $R(8,2)$ computation used $84$ qubits of which $28$ were computational 
qubits. This computation is the largest experimental implementation of a 
scientifically meaningful adiabatic evolution algorithm that has been done to 
date.
\end{abstract}


\maketitle


In recent years first steps have been taken towards experimentally
realizing the computational advantages promised by well-known quantum
algorithms. As with any nascent effort, these initial steps have been limited.
To date the largest experimental implementations of scientifically meaningful
quantum algorithms have used just a handful of qubits. For circuit-based
algorithms,\cite{vander} \textit{seven\/} spin-qubits were used to factor $15$;
while for adiabatic algorithms, \cite{XuZhu} \textit{four\/} spin-qubits
were used to factor $143$. In \textit{both\/} cases compiled versions of the
algorithms were needed to allow factoring with such small numbers of
qubits. Although factoring was the focus of both experiments, other
scientifically significant applications exist.

In Ref.~\onlinecite{Gaitan2011} an algorithm for determining the two-color 
Ramsey numbers was proposed which could be implemented using adiabatic quantum
evolution. Ramsey numbers are part of an active research area in mathematics 
known as Ramsey theory \cite{Graham1990} whose central theme is the emergence
of order in large disordered structures. The disordered structures can be 
represented by an $N$-vertex graph $G$, and the ordered substructures by 
specific graphs $H_{1}$ and $H_{2}$ that are to appear as subgraphs of $G$. 
For two-color Ramsey numbers the subgraphs $H_{1}$ and $H_{2}$ are 
$m$-cliques and $n$-independent sets, respectively. An $m$-clique is a set of 
$m$ vertices that has an edge connecting any two of the $m$ vertices, and an 
$n$-independent set is a set of $n$ vertices in which no two of the $n$ vertices
is joined by an edge. Using Ramsey theory\cite{Graham1990,Bollobas1998}, 
one can prove that a threshold value $R(m,n)$ exists so that for $N \geq 
R(m,n)$, \textit{every\/} graph with $N$ vertices will contain either an 
$m$-clique or an $n$-independent set. The threshold value $R(m,n)$ is an 
example of a two-color Ramsey number. Other types of Ramsey numbers exist, 
though we focus on two-color Ramsey numbers here. Ramsey numbers $R(m,n)$ 
grow extremely quickly and are notoriously difficult to calculate. In fact, for 
$m,n\geq 3$, only nine are presently known\cite{Bollobas1998}.

In the Ramsey number algorithm\cite{Gaitan2011} (RNA) the calculation
of $R(m,n)$ is formulated as an optimization problem which can be solved using
adiabatic quantum evolution.\cite{Farhi2000} Here we present evidence of an
\textit{experimental\/} implementation of the RNA using adiabatic quantum 
evolution and show that it correctly determines the Ramsey numbers $R(3,3)$ 
and $R(m,2)$ for $4\leq m\leq 8$. The experimental computation of $R(8,2)$ 
used a total of $84$ qubits of which $28$ were computational qubits, and
applied an effective interaction coupling 28 qubits. To the best of our
knowledge, this is the largest experimental implementation of a scientifically
meaningful adiabatic evolution algorithm.

We begin this Letter with a brief description of the RNA. We then discuss the
details of experimentally implementing the RNA using adiabatic quantum evolution
on a chip of $106$ superconducting flux qubits, and follow with a presentation 
of our experimental results. Finally, we close with a discussion of what has 
been found.

\textbf{1.~Ramsey number algorithm:}
We briefly describe the construction of the RNA;
see Ref.~\onlinecite{Gaitan2011} and the Supplementary Information (SI) for
a detailed presentation. 

As computation of $R(m,n)$ is intimately connected with the presence/absence
of edges, we associate a bit-variable $a_{v,v'}$ with each pair of vertices
$(v,v')$ in an $N$-vertex graph $G$, and set $a_{v,v'}=1$ ($0$) when $v$ and
$v'$ are (are not) joined by an edge in $G$. There are thus
$L_{N}=\binom{N}{2}\equiv N(N-1)/2$ bit-variables which we collect into the
bit-vector (bit-string) $\vc{a} = (a_{2,1}\cdots , a_{N,1},a_{3,2}, \cdots ,
a_{N,2},\cdots , a_{N,N-1})$ of length $L_{N}$. Thus an $N$-vertex graph $G$
determines a unique bit-string $\vc{a}$, and vice versa.
Ref.~\onlinecite{Gaitan2011} (and the SI) showed how to count the number
of $m$-cliques $C_{m}^{N}(\vc{a})$ and $n$-independent sets $I_{n}^{N}(\vc{a})$
in an $N$-vertex graph $G$ using its associated bit-string $\vc{a}$. We can
thus calculate the total number of $m$-cliques and $n$-independent sets
contained in $G$: $h^{N}_{m,n}(\vc{a}) = C^{N}_{m}(\vc{a})+ I^{N}_{n}
(\vc{a})$. Now consider the following combinatorial optimization problem (COP):
For given integers $(N,m,n)$ and cost function $h^N_{m,n}(\vc{a})$ defined as
above, find an $N$-vertex graph $G_{\vc{a}_{\star}}$ that yields the global
minimum of $h^{N}_{m,n}(\vc{a})$. Notice that if $N<R(m,n)$, the global
minimum is $h^N_{m,n}(\vc{a}_{\star})=0$ since Ramsey theory guarantees
that a graph exists that has no $m$-clique or $n$-independent set. Furthermore,
if $N\geq R(m,n)$, Ramsey theory guarantees that 
$h^{N}_{m,n}(\vc{a}_{\star})>0$.

Adiabatic quantum optimization (AQO)\cite{Farhi2000} is a $T=0$ ground-state
method that exploits the adiabatic evolution of a quantum system to solve COPs,
while quantum annealing (QA)\cite{QAPs} is a finite temperature method which 
can also be used to solve COPs even in the presence of decoherence. 
Ref.~\onlinecite{Gaitan2011} described a quantum implementation of the RNA 
using AQO, while in this Letter we present evidence for a QA implementation of
the RNA. (Note that a classical implementation of the RNA is also possible
using a classical optimization algorithm run on a classical computer to solve 
the Ramsey number COP.) Both AQO and QA use the COP cost function 
to define a problem Hamiltonian $H_{P}$ whose ground-state eigenspace contains
all COP solutions. These algorithms evolve the state of a qubit register from 
the ground-state of an initial Hamiltonian $H_{i}$ to a ground-state of $H_{P}$
with high probability in the adiabatic limit. The algorithm dynamics is driven 
by a time-dependent Hamiltonian $H(t) = A(t/t_f) H_{i} + B(t/t_f)H_{P}$, 
where $t_f$ is the algorithm run-time, adiabatic dynamics corresponds to 
$t_f\rightarrow\infty$, and $A(t/t_f)$ ($B(t/t_{f})$) is a positive 
monotonically decreasing (increasing) function with $A(1)=0$ ($B(0)=0$).

In the quantum implementation of the RNA each bit-variable $a_{v,v'}$ is 
promoted to a qubit, thus associating a qubit  with each vertex pair $(v,v')$. 
The bit-strings $\vc{a}$ now label the computational basis (CB) states 
$|\vc{a}\rangle$, and the problem Hamiltonian $H_{P}$ is defined to be diagonal 
in the CB with eigenvalue $h^{N}_{m,n}(\vc{a})$. By construction, the 
ground-state energy of $H_{P}$ vanishes if and only if there is a graph with 
no $m$-cliques or $n$-independent sets. The initial Hamiltonian $H_i$ is the 
standard one for AQO\cite{Farhi2000} and appears in the SI. Its unique 
ground-state is the uniform superposition of all CB states.

Implementation of the RNA using QA computes $R(m,n)$ as follows. First, 
choose $N$ such that $N<R(m,n)$, then run QA on the $L_N$ qubits 
and measure the qubits in the CB at the end of the anneal. This yields a
bit-string $\vc{a}_{\ast}$ which determines the final energy $E=h^{N}_{m,n}
(\vc{a}_{\ast})$.  In the adiabatic limit the result will be $E=0$ since $N<R(m,n)$. 
Now increment $N\rightarrow N+1$, re-run QA on the $L_{N+1}$ qubits, and 
measure the final energy. Repeatedly increment $N$ until $E>0$ first 
occurs, at which point the current value of $N$ will be equal to $R(m,n)$. In 
any real application of the above algorithm the evolution will only be 
approximately adiabatic, and the probability that the measured energy will be 
the ground-state energy will thus be $1-\epsilon$. By running the algorithm 
$k\sim \mathcal{O}(\ln [1-\delta ]/\ln\epsilon )$ times, the probability 
$\delta$ that at least one of the measurement outcomes yields the 
ground-state energy can be made arbitrarily close to $1$.

\textbf{2.~Experimental implementation:}
Our hardware is designed to implement QA using RF-SQUID flux qubits. Each 
qubit is a superconducting loop interrupted by Josephson junctions, and the 
states $|0\rangle$ and $|1\rangle$ correspond to the two directions of 
circulating current about the loop.\cite{Harris2010} The chip hardware uses
Josephson-junction-based devices to produce pairwise
qubit-coupling.\cite{Harris2009} 
By rescaling the chip Hamiltonian by the
inter-qubit coupling energy $J_{AFM}(t) = M_{AFM}|I^{p}_{q}(t)|^{2}$,
the low energy dynamics of the chip can be represented by a quantum Ising
model in a transverse field with Hamiltonian:\cite{Harris3}
\begin{equation*}
H = -A\left( \! \frac{t}{t_f}\!\right) \sum_i  \sigma^x_i + B\left(\!
                           \frac{t}{t_f}\!\right)
          \bigg\{ \!\! \sum_i h_i \sigma^z_i +
\sum_{(i,j)\in E} \hspace{-0.5em} J_{ij} \sigma^z_i \sigma^z_j  \!\bigg\}.
\label{IsingHamEq}
\end{equation*}
Here $M_{AFM}$ is the maximum value of the inter-qubit effective mutual
inductance that the hardware can produce, and $I^{p}_{q}(t)$ is the supercurrent
circulating about the RF-SQUID loop. Although the scale factor $J_{AFM}(t)$ is
time-dependent, its order-of-magnitude can be estimated. For the D-Wave
One device, $M_{AFM} =1.8pH$, and $|I^{p}_{q}(t)| \sim 0.6\mu A$ at the
quantum critical point, so that $J_{AFM}\sim 1GHz$ or $50mK$. With
this rescaling, the local biases  $\{h_i\}$ and coupling strengths $\{J_{ij}\}$ may be
programmed to values in the ranges $[-2,2]$ and $[-1,1]$ respectively, and
the experimentally measured functional forms of the interpolation functions
$A(t/t_f)$ and $B(t/t_f)$ appear in the SI, along with the layout of qubits 
and couplers on the chip.
For further details of the chip hardware, see Ref.~\onlinecite{Johnson2011}.

 The cost function $h^{N}_{m,n} (\vc{a})$ is not yet ready for experimental
implementation for two reasons: (a)~there are $k$-qubit interactions with
$k>2$; and (b)~the qubit couplings do not correspond to the qubit couplings
on the chip. These obstacles are removed as follows.

(\textit{a})~\textit{Reduction to pairwise coupling:}
The SI shows that $C^{N}_m(\vc{a})$ involves interactions coupling
$\binom{m}{2}$ qubits, while $I^{N}_n(\vc{a})$ couples $\binom{n}{2}$ qubits.
These interactions must be reduced to pairwise coupling if $h^{N}_{m,n}(\vc{a})$
is to be realized experimentally. We illustrate how such a reduction can be
achieved by reducing the $3$-bit coupling term $a_{1}a_{2}a_{3}$ to pairwise
coupling using an ancillary bit-variable $b$ and penalty function
$P(a_1,a_2;b) = a_1 a_2 - 2(a_1+a_2)b+3b $.
Notice that $P(a_{1},a_{2};b) = 0$ ($>0$) when the input values for $a_{1}$,
$a_{2}$, and $b$ satisfy $b=a_{1}a_{2}$ ($b\neq a_{1}a_{2}$). Now consider
the quadratic cost function $h(b) = ba_{3}+ \mu P(a_{1}, a_{2};b)$
for given values of $\mu$ and $a_{i}$. For $\mu$ sufficiently large,
$h(b)$ is minimized when $b$ satisfies the equality constraint $b=a_{1}a_{2}$
which causes the penalty function to vanish. The optimum cost is then
$h(b=a_{1}a_{2}) = a_{1}a_{2}a_{3}$ which reproduces the $3$-bit coupling
term using a \textit{quadratic\/} cost function. This example is generalized in
the SI to produce the quadratic cost function used to calculate $R(m,2)$.

(\textit{b})~\textit{Matching spin to qubit connectivity:}
A cost function with only pairwise qubit coupling may still not be experimentally
realizable as the qubit couplings needed may not match the qubit couplings
available on chip. The primal graph (PG) of a quadratic cost function is
the graph whose vertices are the qubit variables, and whose edges indicate
pairwise-coupled qubits. An arbitrary PG can be embedded into a
sufficiently large qubit graph having the qubit layout and connectivity present
on the chip. An embedding maps a PG vertex to one or more vertices in the
qubit graph, where the image vertices form a connected subgraph of the qubit
graph. We link this connected set of qubits together with strong ferromagnetic
couplings 
of strength $\lambda$ so that in the lowest energy state these qubits have 
identical Bloch vectors. An example of this ferromagnetic coupling procedure is 
given in the SI.


\textbf{3.~Embedding Ramsey problems:}
We examined all Ramsey number COPs that could be solved using the 106 qubits
available on the chip. Specifically, $R(m,2)$ with $4\leq m\leq 8$, and
$R(3,3)$. Here we describe the embedding of these problems on to the chip.

(\textit{a})~$R(m,2)$: Since an $N$-vertex graph $G_\vc{a}$ with
$N<m$ cannot contain an $m$-clique, it follows that $C^{N}_{m}(\vc{a})=0$
for all such $G_{\vc{a}}$. Thus, for $N<m$, $h^{N}_{m,2}(\vc{a}) = I^{N}_{2}
(\vc{a})= \overline{a}_{1}+\cdots +\overline{a}_{L_{N}}$, where
$\overline{a}_{i} =1 - a_{i}$. This produces a problem Hamiltonian $H_{P}$
with $L_{N}$ \textit{uncoupled\/} qubits which is easily mapped onto the
chip. Now consider $N=m$. Defining $L=L_{m} = \binom{m}{2}$, we have
$C^m_m(\vc{a}) = a_1 a_2 \cdots a_{L-1} a_{L}$, and $h^m_{m,2}(\vc{a})
=C^m_m(\vc{a}) + I^m_2(\vc{a})$. The $L$-bit interaction in $C^{m}_{m}
(\vc{a})$ is reduced to pairwise coupling by introducing: (i)~ancillary bit
variables $b_{2},\cdots , b_{L-1}$, and (ii)~imposing the constraints $b_{L-1}
=a_{L-1}a_{L}$ and $b_{j}=a_{j}b_{j+1}$ ($j=2,\cdots ,L-2$) through the
penalty function $P(\vc{a};\vc{b}) = P(a_{L-1},a_L;b_{L-1})+ \sum_{j=2}^{L-2}
P(a_j, b_{j+1};b_j)$, where $P(a,b;c)$ was defined in Section~2(a). 
The $R(m,2)$ cost function for $N=m$ is then
$
h^m_{m,2}(\vc{a},\vc{b}) = \left\{ a_1 b_2 + \mu\, P(\vc{a};\vc{b})
                                                   \right\}       +I^{m}_{2}(\vc{a}),
$
where $\mu =2$ is the penalty weight value used in all $R(m,2)$ experiments.
Making the substitutions $2\vc{a}=\vc{s}_a+1$ and $2\vc{b}=\vc{s}_b+1$
expresses the cost function in terms of Ising spin variables $\vc{s}_{a}$ and
$\vc{s}_{b}$. The PG for the pairwise interactions present in
$h^m_{m,2}(\vc{a},\vc{b})$ appears in the SI. We have
embedded this PG into the hardware up to $N=m=8$. In the SI we display the
embedding that was used to determine $R(8,2)$ which used 28 computational
qubits, 26 ancilla qubits to reduce interactions to pairwise, and 30 qubits to
match the PG connectivity to the qubit connectivity available on the chip
for a total of 84 qubits.

(\textit{b})~$R(3,3)$:
We also determined $R(3,3)$ by examining $N=4,5,6$. The cost functions
for these cases are:
\begin{align*}
h^4_{3,3}(\vc{a}) &= f_{1,2,4} + f_{1,3,5} + f_{2,3,6} + f_{4,5,6}; \\
h^5_{3,3}(\vc{a}) &= f_{1,2,5} + f_{1,3,6} + f_{1,4,7} + f_{2,3,8} +
f_{2,4,9} +  \\ &\phantom{=} \;\; f_{3,4,10} + f_{5,6,8} + f_{5,7,9} +
f_{6,7,10} + f_{8,9,10} ; \\
h^6_{3,3}(\vc{a}) &= f_{1,2,6} + f_{1,3,7} + f_{1,4,8} + f_{1,5,9} +
f_{2,3,10} + f_{2,4,11} \\ &\phantom{=} \;\; +f_{2,5,12} + f_{3,4,13}
+ f_{3,5,14} + f_{4,5,15} + f_{6,7,10}\\ &\phantom{=}\;\; + f_{6,8,11}
+ f_{6,9,12} + f_{7,8,13} + f_{7,9,14} + f_{8,9,15}\\ &\phantom{=} \;\;
+ f_{10,11,13} + f_{10,12,14} + f_{11,12,15} + f_{13,14,15} ;
\end{align*}
where $f_{i,j,k} = a_i a_j a_k + \overline{a}_i \overline{a}_j \overline{a}_k$.
Notice that $f_{i,j,k}$ can be re-written as $f_{i,j,k} = -2 + \overline{a}_i +
\overline{a}_j+\overline{a}_k + a_i a_j + a_i a_k + a_j a_k$, which only
contains \textit{pairwise\/} couplings, making ancillary $\vc{b}$-qubits
unnecessary. The largest of these problems is for $N=6$ whose PG has 15
vertices and 60 edges.
We can reduce its size slightly by exploiting the identity $h^N_{m,n}(\vc{a})
= h^N_{n,m}(\overline{\vc{a}})$. For $m=n$ this yields a two-fold symmetry:
if $\vc{a}_\star$ is a global minimum of $h^N_{m,m}$, so is
$\overline{\vc{a}}_\star$. Thus, we can fix one variable (say $a_1=0$) and
optimize over the remaining variables $\vc{a}'$. Optimal solutions
then have the form $(0,\vc{a}'_\star)$ and $(1,\overline{\vc{a}}'_\star)$.
With this simplification, the PG of $h_{3,3}^6(0,\vc{a}')$ has 14 vertices
and 52 edges. We show its embedding into the chip hardware in the SI.

\textbf{4.~Results:}
To solve a given  Ramsey COP specified by the parameters $\vc{h}$ 
and $\vc{J}$, the chip must first be programmed to fix these values in hardware.
For the largest problem we solved ($R(8,2)$ using 84 qubits) this took roughly
270 ms. After programming we iterate many cycles of annealing and readout.
Each annealing cycle has duration $t_f=1$ ms, and readout of the qubits takes
1.5 ms per sample. Programming only occurs once so the total runtime required
to obtain $S\/$ Ramsey output samples is $270+(1+1.5)S$ ms. As the hardware
is an analog device, there is limited precision to which $\vc{h}$ and $\vc{J}$
can be specified. For a COP whose ground state is sensitive to parameter
settings this could pose serious difficulties. However, the Ramsey COP 
requires
specification of only a few distinct integral values, and ground states are quite
stable to parameter perturbations. The reader is referred to the SI for further
discussion of: (i)~the quantum annealing rate and the distribution of final
energies; (ii)~the experimental temperature; and (iii)~parameter noise.

\begin{figure}[h!]
\begin{center}
\mbox{\subfigure[$\; N=7$]{\includegraphics
[trim= 0 6cm 0 7cm, clip, width=8cm]{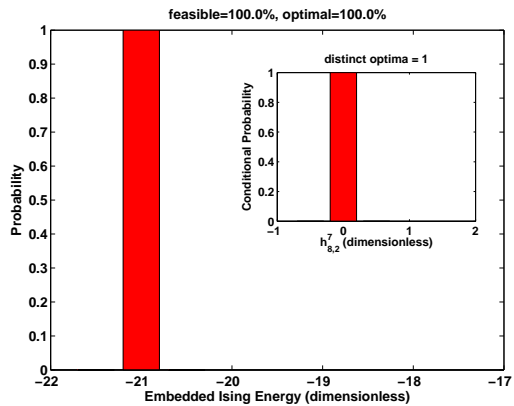}}}
\mbox{\subfigure[$\; N=8$]{\includegraphics
[trim= 0 6cm 0 7cm, clip, width=8.0cm]{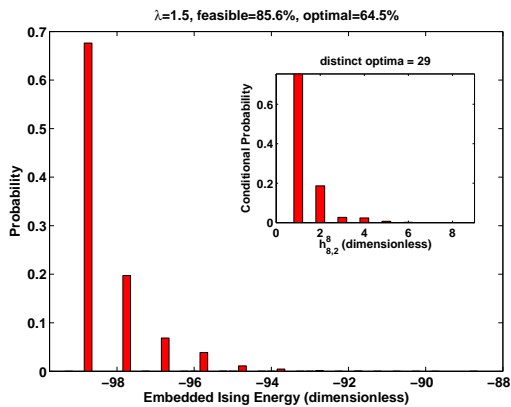}} }
\end{center}
\caption{\textbf{Energy histograms for \boldmath{$R(8,2)$}} on graphs below
($N=7$) and at ($N=8$) the Ramsey transition\vspace{-0.15in}.}\label{R288Fig}
\end{figure}

\begin{figure}[h!]
\begin{center}
\mbox{\subfigure[$\; N=5$]{\includegraphics
[trim= 0 6cm 0 7cm, clip, width=8.0cm]{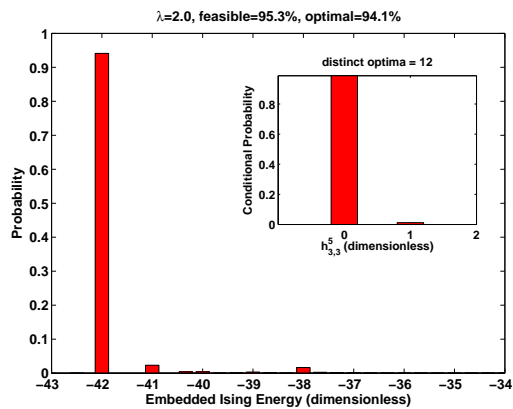}} }
\mbox{\subfigure[$\; N=6$]{\includegraphics
[trim= 0 6cm 0 7cm, clip, width=8.0cm]{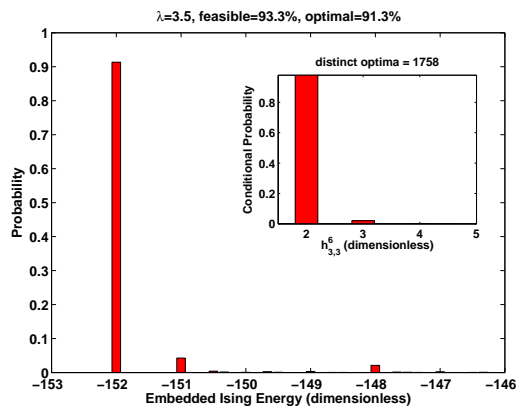}} }
\end{center}
\caption{\textbf{Energy histograms for \boldmath{$R(3,3)$}} on graphs below
($N=5$) and at ($N=6$) the Ramsey transition\vspace{-0.in}.}\label{R33Fig}
\end{figure}

Figs.~\ref{R288Fig}
 \hspace{-0.2em} and \ref{R33Fig} present our results for $R(8,2)$ and $R(3,3)$.
Due to space limitations, our full set of results ($R(3,3)$ and $R(m,2)$
with $4\leq m \leq 8$) appear in the SI, though Table~\ref{table1}
\begin{table*}[hbt!]
\caption{\label{table1} Results for Ramsey numbers $R(m,2)=m$ for
$4\leq m\leq 8$, and $R(3,3)=6$. Here $N$ is the number of graph vertices;
$E_{gs}$ and $D$ are the ground-state energy and degeneracy, respectively,
for the problem Hamiltonian $H_{P}$; and for each Ramsey number, the
experimental results are followed by the theoretical predictions from
Ref.~\onlinecite{Gaitan2011} in parenthesis.}
\begin{ruledtabular}
\begin{tabular}{|c|c|c||c|c|c||c|c|c||c|c|c||c|c|c||c|c|c| }
\multicolumn{3}{|c||}{$\mathbf{R(2,4)}$} &
  \multicolumn{3}{c||}{$\mathbf{R(2,5)}$} &
    \multicolumn{3}{c||}{$\mathbf{R(2,6)}$} &
     \multicolumn{3}{c|}{$\mathbf{R(2,7)}$} &
    \multicolumn{3}{c||}{$\mathbf{R(2,8)}$} &
     \multicolumn{3}{c|}{$\mathbf{R(3,3)}$}\\\hline
$N$ & $E_{gs}$ & $D$ &
$N$ & $E_{gs}$ & $D$ &
$N$ & $E_{gs}$ & $D$ &
$N$ & $E_{gs}$ & $D$ &
$N$ & $E_{gs}$ & $D$ &
$N$ & $E_{gs}$ & $D$ \\\hline
$3$ & $0$($0$) & $1$($1$) &
$4$ & $0$($0$) & $1$($1$) &
$5$ & $0$($0$) & $1$($1$) &
$6$ & $0$($0$) & $1$($1$) &
$7$ & $0$($0$) & $1$($1$) &
$5$ & $0$($0$) & $12$($12$)\\
$4$ & $1$($1$) & $7$($7$) &
$5$ & $1$($1$) & $11$($11$) &
$6$ & $1$($1$) & $16$($16$) &
$7$ & $1$($1$) & $22$($22$) &
$8$ & $1$($1$) & $29$($29$) &
$6$ & $2$($2$) & $1758$($1760$)\\
\end{tabular}
\end{ruledtabular}
\end{table*}
contains a summary of all results, along with corresponding theoretical
predictions. Both Figures~\ref{R288Fig} and \ref{R33Fig} display histograms
that plot the relative-frequency of energy values obtained by programming
the chip and running $10^5$ annealing and readout cycles, yielding $10^5$
$\vc{s}$-spin configurations. In the main figures, histograms
of the energies of the Ising problem sent to the hardware are plotted. These
Ising cost functions include the Ramsey cost function $h^N_{m,n}(\vc{s})$,
and the ferromagnetic penalties, 
$\lambda$, enforcing the equality constraints amongst
qubits that represent the same PG spin variable. The ferromagnetic
penalty weight used for embedding was adjusted so that at least 85\% of output
$\vc{s}$-configurations satisfied the equality constraints. These feasible spin
configurations were translated back to the original $\vc{a}$ variables
and the cost/energy function $h^N_{m,n}(\vc{a})$ evaluated.
The resulting energy values were binned and plotted in the inset histograms.
Further discussion of the equality constraint protocol appears in the SI.

\begin{figure}[h!]
\begin{center}
\includegraphics
[trim= 2cm 8.3cm 0cm 8cm, clip, width=10cm]{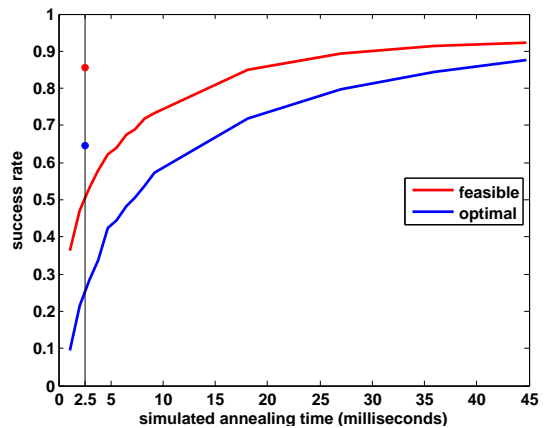}
\end{center}
\caption{Plot of simulated annealing (SA) success probability to determine 
optimal and feasible spin-configurations versus run-time for $R(8,2)$ at 
$N=8$. The SA cooling schedule is exponential; and the initial and final 
temperatures were optimized for maximum success probability. For comparison,
the D-Wave hardware results are also shown for a runtime of $2.5ms$.
See text for further discussion\vspace{-0.17in}.}
\label{saplot}
\end{figure}

Fig.~\ref{R288Fig} presents our results for $R(8,2)$ for $N=7$ and $8$,
while $N=6$ appears in the SI. The $R(8,2)$ experiment was the
largest of the $R(m,2)$ problems considered. Of the $\vc{s}$-configurations
returned by the hardware for $N=8$, approximately 65\% are global minima
of $h^{8}_{8,2}$. By comparison, steepest-descent local-search started from
a random spin configuration finds a globally minimal configuration with less than
0.01\% probability. 
Note that classical/thermal annealing can be ruled out as the source of 
optimization efficacy. First, it is clear that the hardware is not realizing classical 
annealing since the final distribution of low energy states is not Boltzmann 
distributed as discussed in the SI, Sec.~VI~B, and furthermore, the 
temperature of the hardware is never varied during the experiments. Finally, 
we compare the optimization efficacy of the hardware with that of an 
efficient C-implementation of simulated annealing that was run on a standard 
$8$Gb, $2.66$GHz desktop computer. The results of Fig.~\ref{saplot} 
show that at a runtime of $2.5ms$ (which is the $1ms$ runtime plus $1.5ms$ 
read-out time of the Ramsey number experiments), the hardware obtains 
significantly higher success rates for finding both feasible and optimal final 
spin-configurations than does simulated annealing. Returning to 
Fig.~\ref{R288Fig}, examination of the inset histograms for $N=7$ ($8$) we 
see that:
(i)~$h_{min}= 0\; (1)$; (ii)~the probability for $h=0\; (1)$ is approximately
$1.0\; (0.65)$; and (iii)~the number of optimal $\vc{a}$-configurations/graphs
is $1$ ($29$). The reader is referred to the SI for an explanation of how the
probability for an optimal spin-configuration is determined. The energies 
$h_{min}$ found for $N=7$ and $8$ agree with the final ground-state (GS) 
energies found in Ref.~\onlinecite{Gaitan2011}, indicating that the hardware 
finds 
the final GS with high-probability. As $h_{min}$ jumps from $0\rightarrow 1$ as 
$N$ goes from $7\rightarrow 8$, the Ramsey protocol correctly\cite{Bollobas1998}
identifies $R(8,2) = 8$. Finally, Ref.~\onlinecite{Gaitan2011} showed that the 
number of optimal graphs for $N=7\; (8)$ is $1 \; (29)$ which agrees with what 
was found by the hardware. For $N=7$, the unique optimal $\vc{a}$-configuration 
corresponds to the graph in which every pair of vertices is connected by an edge
and so has no $2$-independent sets or $8$-clique and so has $h^{7}_{8,2}=0$. 
For $N=8$, the $29$ optimal $\vc{a}$-configurations correspond to graphs 
$G_{\vc{a}}$ with $h^{8}_{8,2}(\vc{a})=1$ which are the 28 eight-vertex graphs 
containing a single 2-independent set, and the single 8-vertex graph containing 
an 8-clique.

Fig.~\ref{R33Fig}
shows our results for $R(3,3)$ with $N=5$ and $6$, while $N=4$ appears in the
SI. Together, they show that the
minimum energies for $N=4,5,6$ are $0,0,2$, respectively, and these occur with
probabilities of approximately $93$\%, $94$\%, and $91$\%.
These success rates are much higher than steepest-descent local-search. The
success rate for randomly initialized local search is less than $1$\% for $N=6$.
The energies $h_{min}$ agree with the final GS energies found in 
Ref.~\onlinecite{Gaitan2011} indicating that the hardware again finds the final 
GS with high-probability. As $h_{min}$ jumps from $0\rightarrow 2$ as $N$ goes 
from $5\rightarrow 6$, the Ramsey protocol correctly\cite{Bollobas1998} 
identifies $R(3,3)=6$. Finally, Ref.~\onlinecite{Gaitan2011} showed that the 
number of optimal graphs for $N=4, 5,6$ is $18,12,1760$, respectively, in 
excellent agreement with the hardware results of $18,12,1758$.

\textbf{5.~Discussion:}
We presented results of an experimental implementation of the 
RNA\cite{Gaitan2011}. As the Ramsey numbers found correspond to
known Ramsey numbers, it was possible to validate RNA performance. Agreement
betwen theory and experiment was excellent: experimental implementation of the
RNA correctly determined: (i)~$R(m,2) = m$ for $4\leq m\leq 8$ and $R(3,3) = 
6$; and (ii)~the corresponding final ground-state energies and degeneracies. Our
results provide evidence of a quantum implementation of the RNA based on
adiabatic evolution. Further evidence that the D-Wave hardware implements
quantum annealing has recently been reported\cite{Tetals}. It was argued in
Ref.~\onlinecite{SS} that this evidence did not imply quantum annealing, though
these arguments were refuted in Ref.~\onlinecite{refute}. Finally, we
stress that the optimization formulations necessary for experimental realization
of the RNA were non-trivial. The Ising problems after embedding,
solved  with high success rate by the hardware, have many local minima
which were responsible for the low success rates of iterated local-search. 
In spite of the many local minima, the hardware implementation of the RNA 
correctly determined all of the above Ramsey numbers. The $R(8,2)$ computation
used $84$ qubits of which $28$ were computational qubits, and to
the best of our knowledge is the largest experimental implementation
of a scientifically meaningful adiabatic evolution algorithm.

\noindent\textbf{Acknowledgements} We thank the D-Wave
hardware team for providing this experimental platform, and M. Amin, R.
Harris, T. Lanting, and M. Thom for valuable suggestions. F.~G. thanks T.
Howell~III for continued support.

\end{document}


\title{Supporting Information for ``Experimental determination of
Ramsey numbers''}

\date{\today}

\author{Zhengbing Bian}
\affiliation{D-Wave Systems, Inc., 100-4401 Still Creek Drive,
Burnaby, British Columbia V5C 6G9, Canada}

\author{Fabian Chudak}
\affiliation{D-Wave Systems, Inc., 100-4401 Still Creek Drive,
Burnaby, British Columbia V5C 6G9, Canada}

\author{William G. Macready}
\affiliation{D-Wave Systems, Inc., 100-4401 Still Creek Drive,
Burnaby, British Columbia V5C 6G9, Canada}

\author{Lane Clark}
\affiliation{Department of Mathematics, Southern Illinois University,
Carbondale, IL 62901-4401}

\author{Frank Gaitan}
\affiliation{Laboratory for Physical Sciences, 8050 Greenmead Dr,
College Park, MD 20740}

\begin{abstract}
In this supplement we briefly review the construction of the Ramsey number
quantum algorithm, and discuss its experimental implementation. We then present
the complete set of Ramsey number experimental results, including those that 
could not be included in the manuscript due to space limitations. Next we 
present two examples of embeddings
of the Ramsey problem Hamiltonian $H_{P}$ into the chip hardware. These 
embeddings explicitly show how qubit couplings are laid out on the chip so as 
to reproduce the couplings appearing in the problem Hamiltonian $H_{P}$.
Finally, we close with a discussion of a number of important issues associated
with the Ramsey number experiments and their analysis.
\end{abstract}


\maketitle

The structure of this Supporting Information (SI) is as follows. We begin in
Section~\ref{sec0} with a description of how the Ramsey number quantum
algorithm is constructed, and then discuss its experimental implementation
in Sections~\ref{impcost} and \ref{energyfuncs}. Section~\ref{sec1} 
then presents the complete set of Ramsey number experimental results, 
including those that could not be included in the manuscript due to space 
limitations. Section~\ref{sec1a} 
presents the results for $R(3,3)$, while Section~\ref{sec1b} 
presents the results for the Ramsey numbers $R(m,2)$ with $4\leq m\leq 8$. 
For easy reference, Section~\ref{sec1a} and Section~\ref{sec1b} also include 
the data for $R(3,3)$ with $N=5$ and $6$, and $R(8,2)$ for $N=7$ 
and $8$, respectively, which appear in Section~$4$ of the manuscript. 
Section~\ref{sec2} displays the embedding of the Ramsey energy functions into 
the chip for $R(3,3)$ with $N=6$ (Section~\ref{sec2a}) and $R(8,2)$ with
$N=8$ (Section~\ref{sec2b}). These embeddings represent the most complex
embeddings we encountered in our experimental determination of,
respectively, diagonal ($R(m,m)$) and non-diagonal ($R(m,n)$, $m\neq n$)
Ramsey numbers. Finally, we close in Section~\ref{sec4} with a discussion of 
a number of important issues associated with the Ramsey number experiments
and their analysis.

\section{Ramsey number quantum algorithm}
\label{sec0}

 We briefly describe the Ramsey number quantum
adiabatic algorithm (see Ref.~\onlinecite{G&C2011} for details). We begin by
establishing a 1-1 correspondence between the set of
$N$-vertex graphs and binary strings of length $L=N(N-1)/2$. To each
$N$-vertex graph $G$ there corresponds a unique adjacency matrix $A(G)$
which is an $N\times N$ symmetric matrix with vanishing diagonal matrix
elements, and with off-diagonal element $a_{i,j}=1\, (0)$ when distinct
vertices $i$ and $j$ are (are not) joined by an edge. It follows that $A(G)$
is determined by its lower triangular part. By concatenating column-wise
the matrix elements $a_{i,j}$ appearing below the principal diagonal, we
can construct a unique binary string $g(G)$ of length $L$ for each graph $G$:
\begin{equation}
g(G) \equiv a_{2,1}\cdots a_{N,1}\;a_{3,2}\cdots a_{N,2}\;\cdots \;a_{N,N-1}.
\label{121corr}
\end{equation}

Given the string $g(G)$, the following procedure determines the number of
$m$-cliques in $G$. Choose $m$ vertices, $S_{\alpha} = \{ v_{1},\ldots ,
v_{m}\}$, from the $N$ vertices of $G$ and form the product
$\calC_{\alpha} = \prod_{(v_{j},v_{k}\in S_{\alpha})}^{(j\neq k)}
                                  a_{v_{j},v_{k}} $.
Note that $\calC_{\alpha}=1$ when $S_{\alpha}$ forms an $m$-clique;
otherwise $\calC_{\alpha}=0$. Now repeat
this procedure for all $\rho = \binom{N}{m}$
ways of choosing $m$ vertices from $N$ vertices, and form the sum
$\calC^{N}_{m}(G) = \sum_{\alpha =1}^{\rho} \calC_{\alpha}$.
By construction, $\calC^{N}_{m}(G)$ equals the number of $m$-cliques contained 
in $G$. A similar procedure determines the number of $n$-independent sets in
$G$. Briefly, choose $n$ vertices $T_{\alpha} = \{v_{1},\ldots ,v_{n}\}$ from
the $N$ vertices in $G$, and form the product
$\calI_{\alpha} = \prod_{(v_{j},v_{k}\in T_{\alpha})}^{(j\neq k)}
                               \overline{a}_{v_{j},v_{k}}$, where
$\overline{a}_{v_{j},v_{k}} = 1 - a_{v_{j},v_{k}}$. Note that if
$\calI_{\alpha}=1$, then $T_{\alpha}$ forms an $n$-independent set;
otherwise $\calI_{\alpha}=0$. Repeat this for all $\nu = \binom{N}{n}$ ways of
choosing $n$ vertices from $N$ vertices, then form the sum
$\calI^{N}_{n} (G) = \sum_{\alpha =1}^{\nu}\calI_{\alpha}$. By construction,
$\calI^{N}_{m} (G)$ gives the number of $n$-independent sets contained in $G$.
Finally, define
\begin{equation}
h^{N}_{m,n}(G) = \calC^{N}_{m} (G) + \calI^{N}_{n} (G) .
\label{grafcost}
\end{equation}
It follows from the above discussion that $h^{N}_{m,n}(G)$ is the total number 
of $m$-cliques and $n$-independent sets in $G$. Thus $h^{N}_{m,n}(G) \geq 0$ 
for all graphs $G$; and $h^{N}_{m,n}(G) = 0$ if and only if $G$ does not contain
an $m$-clique or $n$-independent set.

We use $h^{N}_{m,n}(G)$ as the cost function for the following
combinatorial optimization problem. For given integers $N$, $m$ and $n$, and
with $h^{N}_{m,n}(G)$ defined as above, find an $N$-vertex graph $G_{\ast}$
that yields the global minimum of $h^{N}_{m,n}(G)$. Notice that if $N<R(m,n)$, 
the (global) minimum will be $h^{N}_{m,n}(G_{\ast})=0$ since Ramsey theory 
guarantees that a graph exists which has no $m$-clique or $n$-independent set. 
On the other hand, if $N\geq R(m,n)$, Ramsey theory guarantees $h^{N}_{m,n}
(G_{\ast})>0$. If we begin with $N<R(m,n)$ and increment $N$ by $1$ until we 
first find $h^{N}_{m,n}(G_{\ast})> 0$, then the corresponding $N$ will be 
exactly $R(m,n)$. We now show how this combinatorial optimization problem 
can be solved using adiabatic quantum evolution, which then becomes the basis 
for a quantum algorithm to compute $R(m,n)$.

\begin{figure}
\includegraphics[width=10cm]{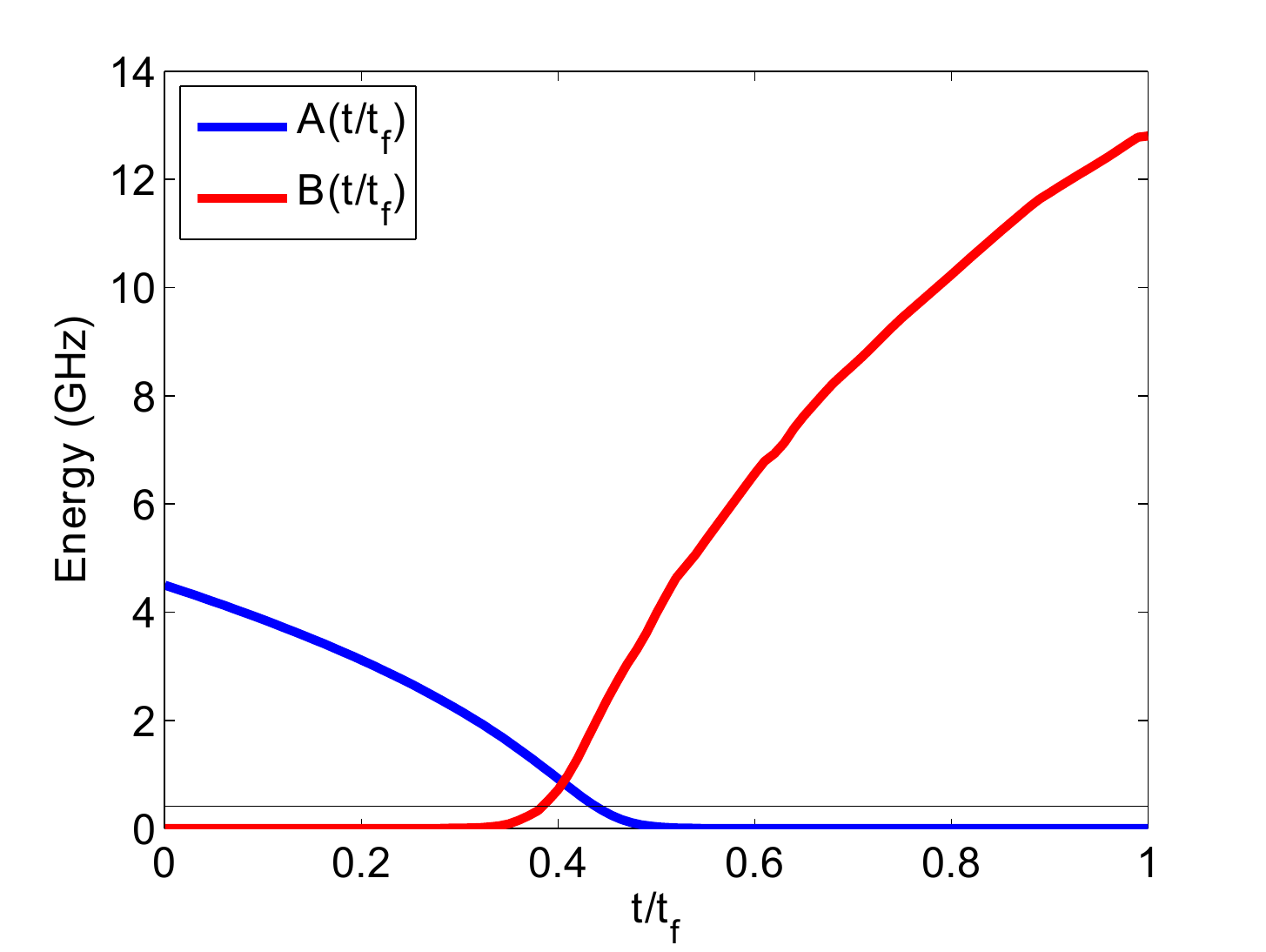}
\caption{\textbf{Interpolation functions $A(t/t_f)$ and $B(t/t_f)$.}
The functional forms of the experimentally measured interpolation functions
$A(t/t_{f})$ and $B(t/t_{f})$ are shown. The QA run-time $t_f$ can be
adjusted for times ranging from $20$--$20\,000\;\mu$s. For comparison, the
temperature at which the experiment is performed is shown as the line at
roughly $0.4$ GHz.} \label{interpFig}
\end{figure}

The adiabatic quantum evolution (AQE)
algorithm \cite{Farhi2000}
exploits the adiabatic dynamics of a quantum system to solve combinatorial
optimization problems. The AQE algorithm uses the optimization problem cost
function to define a problem Hamiltonian $H_{P}$ whose ground-state
eigenspace encodes all problem solutions. The algorithm evolves the state of
an $L$-qubit register from the ground-state of an initial Hamiltonian $H_{i}$
to the ground-state of $H_{P}$ with probability approaching $1$ in the
adiabatic limit. An appropriate measurement at the end of the adiabatic
evolution yields a solution of the optimization problem almost certainly. The
time-dependent Hamiltonian $H(t)$ for local AQE is
\begin{equation}
H(t) = A(t/t_{f})H_{i} + B(t/t_{f}) H_{P},
\label{aqeHam}
\end{equation}
where $t_{f}$ is the algorithm run-time; adiabatic dynamics corresponds to
$t_{f}\rightarrow \infty$; and $A(t/t_f)$ ($B(t/t_{f})$) is a positive monotonically
decreasing (increasing) function with $A(1)=0$ ($B(0)=0$).
The experimentally measured functional forms of the interpolation functions
$A(t/t_f)$ and $B(t/t_f)$
are shown in Fig.~\ref{interpFig}.

To map the optimization problem associated with computing $R(m,n)$ onto
an adiabatic quantum computation, we begin with the 1-1 correspondence
between $N$-vertex graphs $G$ and length $L=N(N-1)/2$ binary strings $g(G)$.
From Eq.~(\ref{121corr}) we see that position along the string is indexed by
the graph edges ($i,j$). We thus identify a qubit with each vertex pair
($i,j$), and will thus need $L$ qubits. Defining the computational basis states
(CBS) to be the eigenstates of $\sigma_{z}^{0}\otimes\cdots\otimes
\sigma_{z}^{L-1}$, we identify the $2^{L}$ graph strings $g(G)$ with the
$2^{L}$ CBS: $g(G)\rightarrow |g(G)\rangle$. The problem Hamiltonian $H_{P}$ is 
defined to be diagonal in the computational basis with eigenvalue $h^{N}_{m,n}
(G)$ associated with eigenstate $|g(G)\rangle$:
\begin{equation}
H_{P}|g(G)\rangle = h^{N}_{m,n}(G)|g(G)\rangle .
\label{HPdef}
\end{equation}
Note that the ground-state energy of $H_{P}$ will be zero iff there is a graph
with no $m$-cliques or $n$-independent sets. An operator expression for $H_{P}$
appears in Ref.~\onlinecite{G&C2011}. The initial Hamiltonian $H_{i}$ is chosen
to be
\begin{equation}
H_{i} = -\sum_{i=0}^{L-1}\sigma_{x}^{i} ,
\label{Hidef}
\end{equation}
where $I^{i}$ and $\sigma_{x}^{i}$ are, respectively, the identity and x-Pauli
operator for qubit $i$. The ground-state of $H_{i}$ is the easily constructed
uniform superposition of CBS\cite{Farhi2000}.

The quantum algorithm for computing $R(m,n)$ begins by setting $N$ equal
to a strict lower bound for $R(m,n)$ which can be found using the probabilistic
method \cite{spenc} or a table of two-color Ramsey numbers \cite{Bollobas}.
The AQE algorithm is run on $L_{N}=N(N-1)/2$ qubits, and the energy $E$ is
measured at the end of algorithm execution. In the adiabatic limit the result
will be $E=0$ since $N<R(m,n)$. The value of $N$ is incremented
$N\rightarrow N+1$, the AQE algorithm is re-run on $L_{N+1}$ qubits, and the
energy is measured at the end of algorithm execution. This process is repeated
until $E>0$ first occurs, at which point the current value of $N$ will be equal
to the $R(m,n)$. Note that any real application of AQE will only be
approximately adiabatic. Thus the probability that the measured energy $E$
will be the ground-state energy will be $1-\epsilon$. In this case, the
algorithm must be run $k\sim\mathcal{O}(\ln [1-\delta ]/\ln\epsilon)$ times
so that, with probability $\delta > 1-\epsilon$, at least one of the measurement
outcomes will be the true ground-state energy. We can make $\delta$ arbitrarily
close to $1$ by choosing $k$ sufficiently large.

Although most discussions of QA assume zero-temperature, all experiments here
were performed at 20 mK, or roughly 0.4 GHz. For comparison, peak values of 
the interpolation functions $A(t/t_{f})$ and $B(t/t_{f})$ are of order 10 GHz 
(see Fig.~\ref{interpFig}). Refs.~\onlinecite{Amin2008} and 
\onlinecite{Amin2009} showed that finite temperature need not 
destroy the efficacy of QA. In our experiments we select the lowest energy 
configuration observed over many annealing cycles to compensate for the 
stochastic influence of non-zero temperature.

\section{Experimental implementation of Ramsey cost functions}
\label{impcost}

\begin{figure}
\includegraphics[width=8cm]{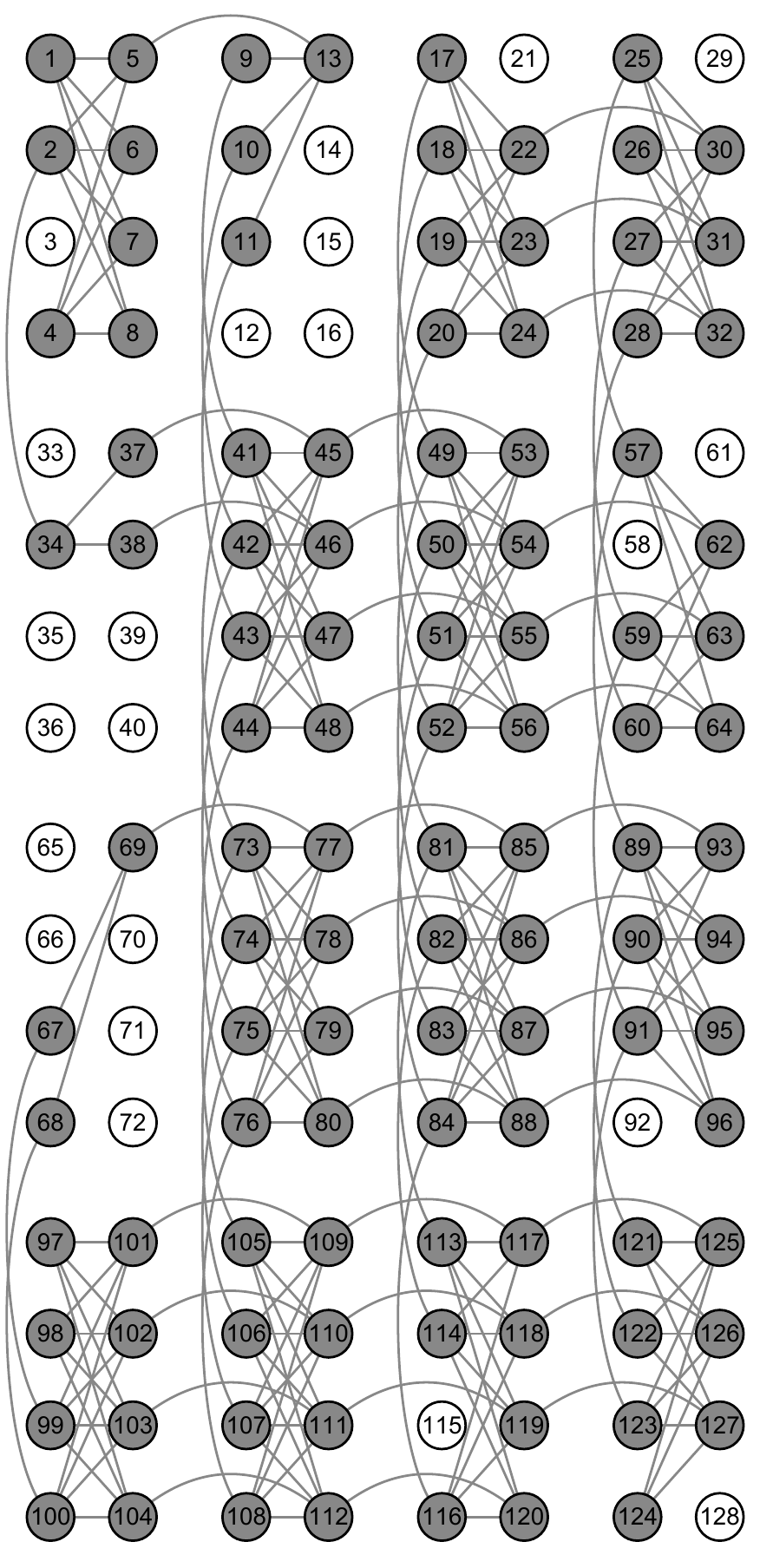}
\caption{\textbf{Layout of qubits and couplers.}
The chip architecture is a $4\times 4$ array of unit cells, with each unit cell
containing 8 qubits. Within a unit cell, each of the $4$ qubits in the left-hand
partition (LHP) connects to all $4$ qubits in the right-hand partition (RHP),
and vice versa. A qubit in the LHP (RHP) of a unit cell also connects to the
corresponding qubit in the LHP (RHP) in the units cells above and below
(to the left and right of) it. Most qubits couple to 6 neighbors. Qubits are
labeled from 1 to 128, and edges between qubits indicate couplers which may
take programmable values. Grey qubits indicate usable qubits, while white
qubits indicate qubits which, due to fabrication defects, could not be
calibrated to operating tolerances and were not used. All experiments
were done on a chip with 106 usable qubits.
} \label{qubitFig}
\end{figure}

To match the notation used in the manuscript we make the substitution $g(G)
\rightarrow \vc{a}$ in Eq.~(\ref{121corr}) and write the Ramsey cost
function as $h^{N}_{m,n}(\vc{a})$. The cost function $h^{N}_{m,n}
(\vc{a})$ is not yet ready for experimental implementation for two reasons: 
(a)~there are $k$-qubit interactions with $k>2$; and (b)~the qubit couplings
do not correspond to the qubit couplings in Fig.~\ref{qubitFig}. These 
obstacles are removed as follows.

(\textit{a})~\textit{Reduction to pairwise coupling:}
Section~\ref{sec0} above showed that $C^{N}_m(\vc{a})$ involves interactions 
coupling
$\binom{m}{2}$ qubits, while $I^{N}_n(\vc{a})$ couples $\binom{n}{2}$ qubits.
These interactions must be reduced to pairwise coupling if $h^{N}_{m,n}(\vc{a})$
is to be realized experimentally. We illustrate how such a reduction can be
achieved by reducing the $3$-bit coupling term $a_{1}a_{2}a_{3}$ to pairwise
coupling using an ancillary bit-variable $b$ and penalty function
\begin{equation}
P(a_1,a_2;b) = a_1 a_2 - 2(a_1+a_2)b+3b . \label{penFuncEq}
\end{equation}
Notice that $P(a_{1},a_{2};b) = 0$ ($>0$) when the input values for $a_{1}$,
$a_{2}$, and $b$ satisfy $b=a_{1}a_{2}$ ($b\neq a_{1}a_{2}$). Now consider
the quadratic cost function $h(b) = ba_{3}+ \mu P(a_{1}, a_{2};b)$
for given values of $\mu$ and $a_{i}$. For $\mu$ sufficiently large,
$h(b)$ is minimized when $b$ satisfies the equality constraint $b=a_{1}a_{2}$
which causes the penalty function to vanish. The optimum cost is then
$h(b=a_{1}a_{2}) = a_{1}a_{2}a_{3}$ which reproduces the $3$-bit coupling
term using a \textit{quadratic\/} cost function. This example is generalized in
Sec.~$3$ to produce the quadratic cost function used to calculate
$R(m,2)$.

(\textit{b})~\textit{Matching spin to qubit connectivity:}
A cost function with only pairwise qubit coupling may still not be experimentally
realizable as the qubit couplings needed may not match the qubit couplings
available on chip. The primal graph (PG) of a quadratic cost function is
the graph whose vertices are the qubit variables, and whose edges indicate
pairwise-coupled qubits. An arbitrary PG can be embedded into a
sufficiently large qubit graph having the structure of Fig.~\ref{qubitFig}. 
An
embedding maps a PG vertex to one or more vertices in the qubit graph, where
the image vertices form a connected subgraph of the qubit graph. We link this
connected set of qubits together with strong ferromagnetic couplings so that
in the lowest energy state these qubits have identical Bloch vectors. For
example, to couple qubits 104 and 75  in Fig.~\ref{qubitFig} (which are not
directly coupled) with coupling strength $\mathcal{J}$, we ferromagnetically
couple qubits 104, 112, and 107 using strongly negative $J_{104,112}$ and
$J_{107,112}$ values. The desired coupling is then carried along the edge
connecting qubits 107 and 75 with $J_{75,107} = \mathcal{J}$.

\section{Ramsey energy functions}
\label{energyfuncs}

We examined a number of Ramsey problems which could be solved using the 106
qubits available in Fig.~\ref{qubitFig}. Specifically, $R(m,2)$ with $4\leq m
\leq 8$ and $R(3,3)$.

(\textit{a})~$R(m,2)$: Since an $N$-vertex graph $G_\vc{a}$ with
$N<m$ cannot contain an $m$-clique, it follows that $C^{N}_{m}(\vc{a})=0$
for all such $G_{\vc{a}}$. Thus, for $N<m$, $h^{N}_{m,2}(\vc{a}) = I^{N}_{2}
(\vc{a})= \overline{a}_{1}+\cdots +\overline{a}_{L_{N}}$, where
$\overline{a}_{i} =1 - a_{i}$. This produces a problem Hamiltonian $H_{P}$
with $L_{N}$ \textit{uncoupled\/} qubits which is easily mapped onto the
chip. Now consider $N=m$. Defining $L=L_{m} = \binom{m}{2}$, we have
$C^m_m(\vc{a}) = a_1 a_2 \cdots a_{L-1} a_{L}$, and $h^m_{m,2}(\vc{a})
=C^m_m(\vc{a}) + I^m_2(\vc{a})$. The $L$-bit interaction in $C^{m}_{m}
(\vc{a})$ is reduced to pairwise coupling by introducing: (i)~ancillary bit
variables $b_{2},\cdots , b_{L-1}$, and (ii)~imposing the constraints $b_{L-1}
=a_{L-1}a_{L}$ and $b_{j}=a_{j}b_{j+1}$ ($j=2,\cdots ,L-2$) through the
penalty function $P(\vc{a};\vc{b}) = P(a_{L-1},a_L;b_{L-1})+ \sum_{j=2}^{L-2}
P(a_j, b_{j+1};b_j)$ (see Eq.~(\ref{penFuncEq})). The $R(m,2)$ cost 
function for $N=m$ is then
$
h^m_{m,2}(\vc{a},\vc{b}) = \left\{ a_1 b_2 + \mu\, P(\vc{a};\vc{b})
                                                   \right\}       +I^{m}_{2}(\vc{a}),
$
where $\mu =2$ is the penalty weight value used in all $R(m,2)$ experiments.
Making the substitutions $2\vc{a}=\vc{s}_a+1$ and $2\vc{b}=\vc{s}_b+1$
expresses the cost function in terms of Ising spin variables $\vc{s}_{a}$ and
$\vc{s}_{b}$. The PG for the pairwise interactions present in 
$h^m_{m,2}(\vc{a},\vc{b})$ appears in Figure~\ref{rm2PrimalFig}. We have 
embedded this PG into the hardware up to $N=m=8$. In Fig.~\ref{r82EmbedFig} 
we display the embedding that was used to determine $R(8,2)$ which used 
28 computational qubits, 26 ancilla qubits to reduce interactions to pairwise, 
and 30 qubits to match the PG connectivity to the qubit connectivity in 
Fig.~\ref{r82EmbedFig} for a total of 84 qubits.

(\textit{b})~$R(3,3)$:
We also determined $R(3,3)$ by examining $N=4,5,6$. The cost functions
for these cases are: 
\begin{align*}
h^4_{3,3}(\vc{a}) &= f_{1,2,4} + f_{1,3,5} + f_{2,3,6} + f_{4,5,6}; \\
h^5_{3,3}(\vc{a}) &= f_{1,2,5} + f_{1,3,6} + f_{1,4,7} + f_{2,3,8} +
f_{2,4,9} +  \\ &\phantom{=} \;\; f_{3,4,10} + f_{5,6,8} + f_{5,7,9} +
f_{6,7,10} + f_{8,9,10} ; \\
h^6_{3,3}(\vc{a}) &= f_{1,2,6} + f_{1,3,7} + f_{1,4,8} + f_{1,5,9} +
f_{2,3,10} + f_{2,4,11} \\ &\phantom{=} \;\; +f_{2,5,12} + f_{3,4,13}
+ f_{3,5,14} + f_{4,5,15} + f_{6,7,10}\\ &\phantom{=}\;\; + f_{6,8,11}
+ f_{6,9,12} + f_{7,8,13} + f_{7,9,14} + f_{8,9,15}\\ &\phantom{=} \;\;
+ f_{10,11,13} + f_{10,12,14} + f_{11,12,15} + f_{13,14,15} ;
\end{align*}
where $f_{i,j,k} = a_i a_j a_k + \overline{a}_i \overline{a}_j \overline{a}_k$.
Notice that $f_{i,j,k}$ can be re-written as $f_{i,j,k} = -2 + \overline{a}_i +
\overline{a}_j+\overline{a}_k + a_i a_j + a_i a_k + a_j a_k$, which only
contains \textit{pairwise\/} couplings, making ancillary $\vc{b}$-qubits
unnecessary. The
largest of these problems is $N=6$ whose PG has 15 vertices and 60 edges.
We can reduce its size slightly by exploiting the identity $h^N_{m,n}(\vc{a})
= h^N_{n,m}(\overline{\vc{a}})$. For $m=n$ this yields a two-fold symmetry:
if $\vc{a}_\star$ is a global minimum of $h^N_{m,m}$, so is
$\overline{\vc{a}}_\star$. Thus, we can fix one variable (say $a_1=0$) and
optimize over the remaining variables $\vc{a}'$. Optimal solutions
then have the form $(0,\vc{a}'_\star)$ and $(1,\overline{\vc{a}}'_\star)$.
With this simplification, the PG of $h_{3,3}^6(0,\vc{a}')$ has 14 vertices
and 52 edges. Fig.~\ref{r33EmbedFig} shows its embedding into the chip hardware.

\section{Complete set of  Ramsey number results}
\label{sec1}

Ref.~\onlinecite{G&C2011} made three predictions that allow theory and
experiment to be compared. The first is that the Ramsey number $R(m,n)$ is
the value of $N$ at which the global minimum $h_{min}$ of the cost function
$h^N_{m,n}(\vc{a})$ first becomes non-zero. The other two are, respectively,
the value of the final ground-state energy $E_{gs} =h_{min} \equiv \min_\vc{a}
h^N_{m,n}(\vc{a})$ and its degeneracy $D$. For reference, Table~\ref{table1}
\begin{table*}
\caption{\label{table1}$\mathrm{Results}^{1}$ for Ramsey numbers $R(m,2)=m$
for $4\leq m\leq 8$ and $R(3,3)=6$.}
\footnotetext{$^{1}N$ is the number of graph vertices;
$E_{gs}$ and $D$ are the ground-state energy and degeneracy,
respectively, for the problem Hamiltonian $H_{P}$; and for each Ramsey number,
the experimental results are followed by the theoretical predictions from
Ref.~\onlinecite{G&C2011} in parenthesis.}
\begin{ruledtabular}
\begin{tabular}{|c|c|c||c|c|c||c|c|c||c|c|c||c|c|c||c|c|c| }
\multicolumn{3}{|c||}{$\mathbf{R(2,4)}$} &
  \multicolumn{3}{c||}{$\mathbf{R(2,5)}$} &
    \multicolumn{3}{c||}{$\mathbf{R(2,6)}$} &
     \multicolumn{3}{c|}{$\mathbf{R(2,7)}$} &
    \multicolumn{3}{c||}{$\mathbf{R(2,8)}$} &
     \multicolumn{3}{c|}{$\mathbf{R(3,3)}$}\\\hline
$N$ & $E_{gs}$ & $D$ &
$N$ & $E_{gs}$ & $D$ &
$N$ & $E_{gs}$ & $D$ &
$N$ & $E_{gs}$ & $D$ &
$N$ & $E_{gs}$ & $D$ &
$N$ & $E_{gs}$ & $D$ \\\hline
$3$ & $0$($0$) & $1$($1$) &
$4$ & $0$($0$) & $1$($1$) &
$5$ & $0$($0$) & $1$($1$) &
$6$ & $0$($0$) & $1$($1$) &
$7$ & $0$($0$) & $1$($1$) &
$5$ & $0$($0$) & $12$($12$)\\
$4$ & $1$($1$) & $7$($7$) &
$5$ & $1$($1$) & $11$($11$) &
$6$ & $1$($1$) & $16$($16$) &
$7$ & $1$($1$) & $22$($22$) &
$8$ & $1$($1$) & $29$($29$) &
$6$ & $2$($2$) & $1758$($1760$)\\
\end{tabular}
\end{ruledtabular}
\end{table*}
from the manuscript is included below which summarizes all our experimental
results and the corresponding theoretical predictions. Examination of the Table
shows that there is excellent agreement between the two.

In Secs.~\ref{sec1a} and \ref{sec1b} we present our results for $R(3,3)$
and $R(m,2)$. For each Ramsey number experiment, histograms of the
final energies are presented. These energies are determined as follows. At the
end of each quantum annealing run, the qubits are measured in the computational
basis, and the outcome is the final spin variable-assignment $\vc{s}$. As
explained in Sec.~$4$ of the manuscript, for those spin configurations
satisfying the equality constraints, the spin configuration $\vc{s}$ is
translated back to the original binary $\vc{a}$ variables. The energy
$E(\vc{a})$ is then calculated from the Ramsey energy function
$h^{N}_{m,n}(\vc{a})$ given in Sec.~$3$ of the manuscript, and its value
entered into the appropriate histogram bin. For each Ramsey number experiment,
energy histograms are presented over a range of $N$ values. As in Sec.~$4$ of
the manuscript, for each $N$, two histograms are given. The main histogram
shows the energy function in terms of spins variable $\vc{s}$, where the
energy function includes the Ramsey energy and the ferromagnetic interactions
enforcing equality constraints between spins representing the same problem
variable. The value of the ferromagnetic coupling strength $\lambda$ used in
the quantum annealing runs appears in the upper left corner of this histogram.
The inset histogram corresponds to the $\vc{a}$-configurations,
and displays the binned Ramsey energies $E(\vc{a})=h^{N}_{m,n}(\vc{a})$.
It is obtained from the spin configurations (of the main histogram) by filtering
out those spin configurations not satisfying the equality constraints as these
are infeasible (in the optimization sense) spin configurations. Each inset
histogram contains the following information: (i)~the set of observed Ramsey
energies $E$, (ii)~the probability (relative frequency) for the energy $E$, and
(iii)~the number of optimal $\vc{a}$-configurations that yielded the minimum
energy $E=h_{min}$.

Having made these preliminary remarks we examine the remaining
Ramsey number results.

\subsection{$\mathbf{R(3,3)=6}$}
\label{sec1a}

\begin{figure}
\begin{center}
\mbox{\subfigure[$\; N=4$]{\includegraphics
[trim= 0 6cm 0 7cm, clip, width=9.5cm]{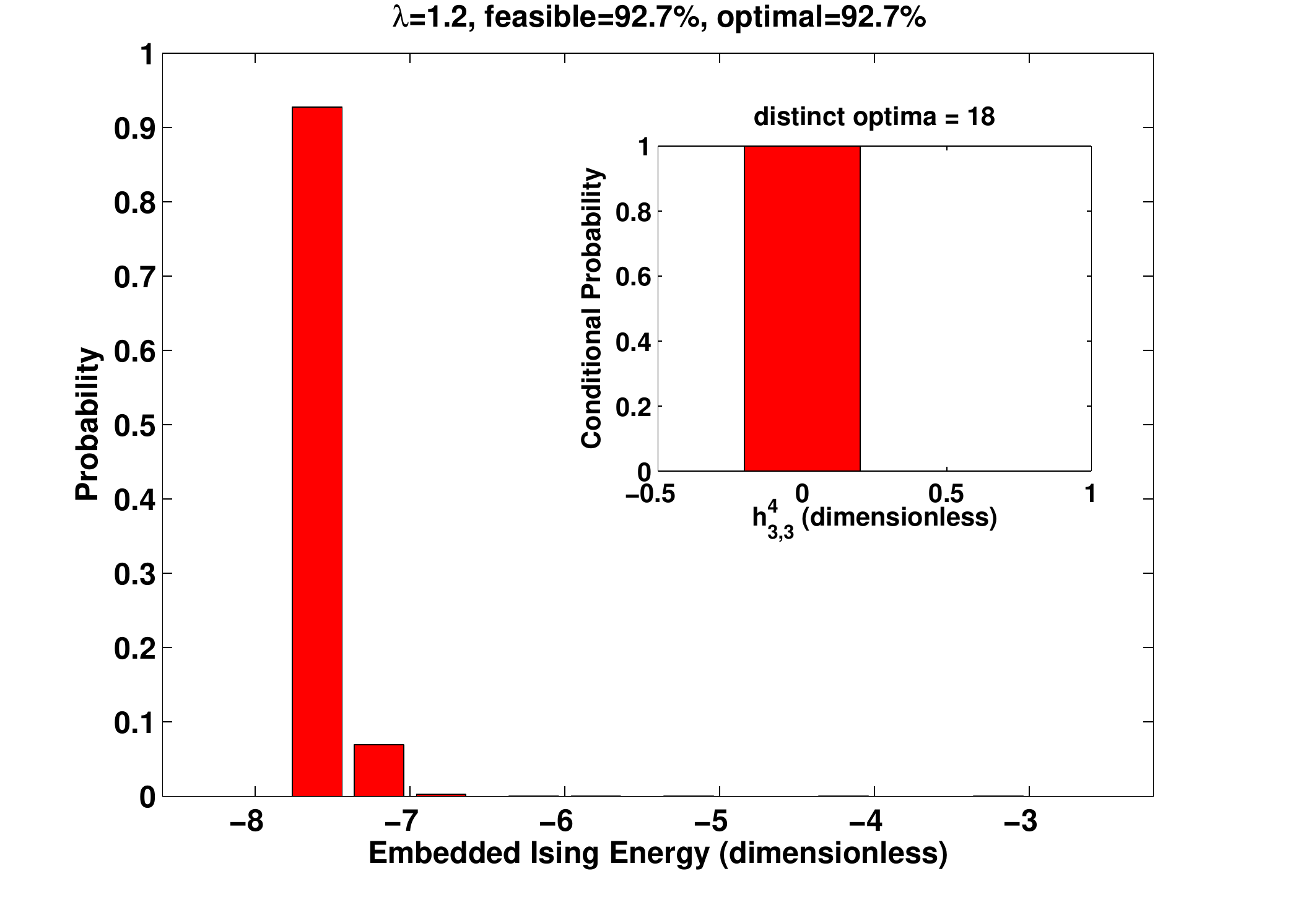} } }
\mbox{\subfigure[$\; N=5$]{\includegraphics
[trim= 0 6cm 0 6cm, clip, width=9.5cm]{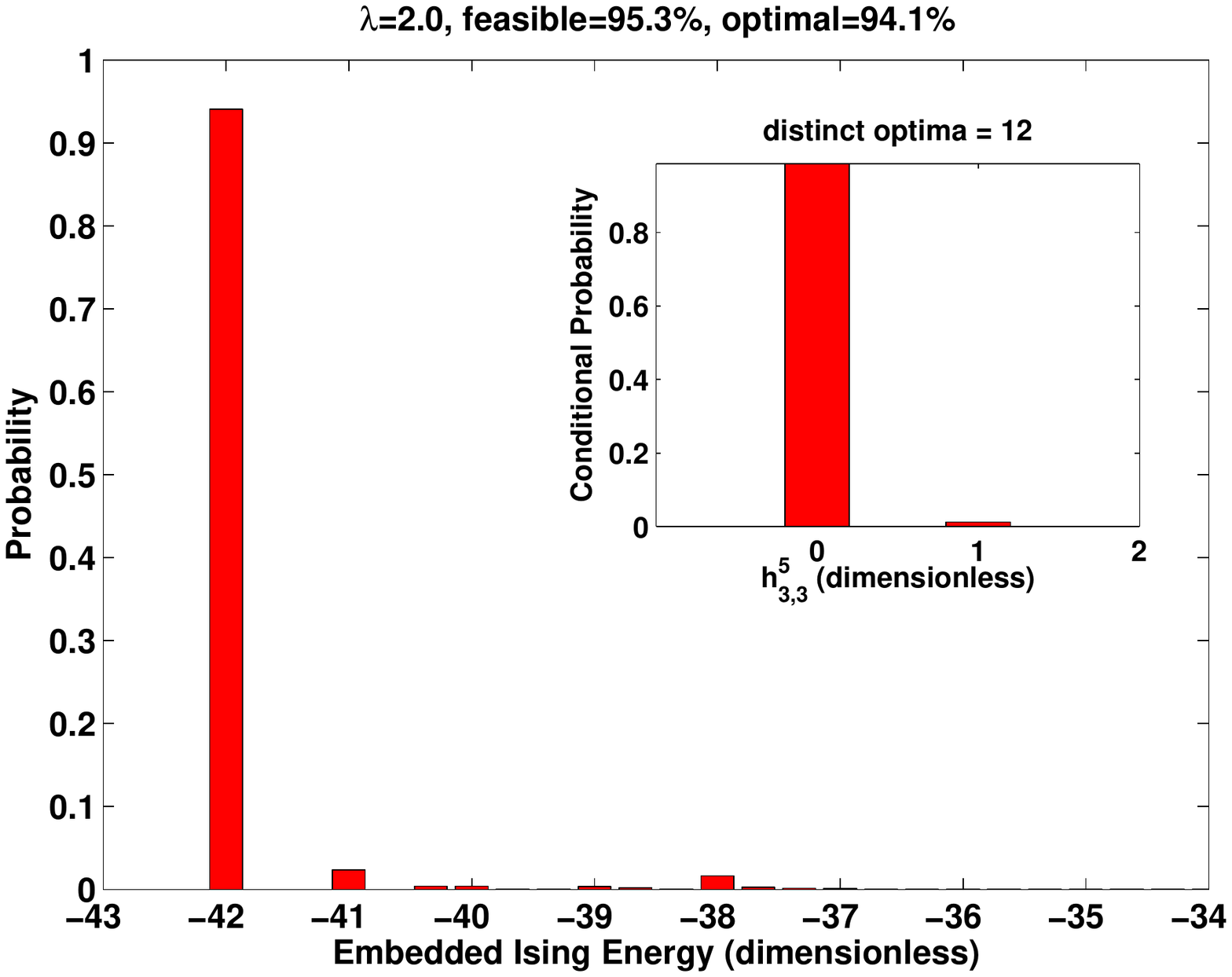} } }
\mbox{\subfigure[$\; N=6$]{\includegraphics
[trim= 0 6cm 0 6cm, clip, width=9.5cm]{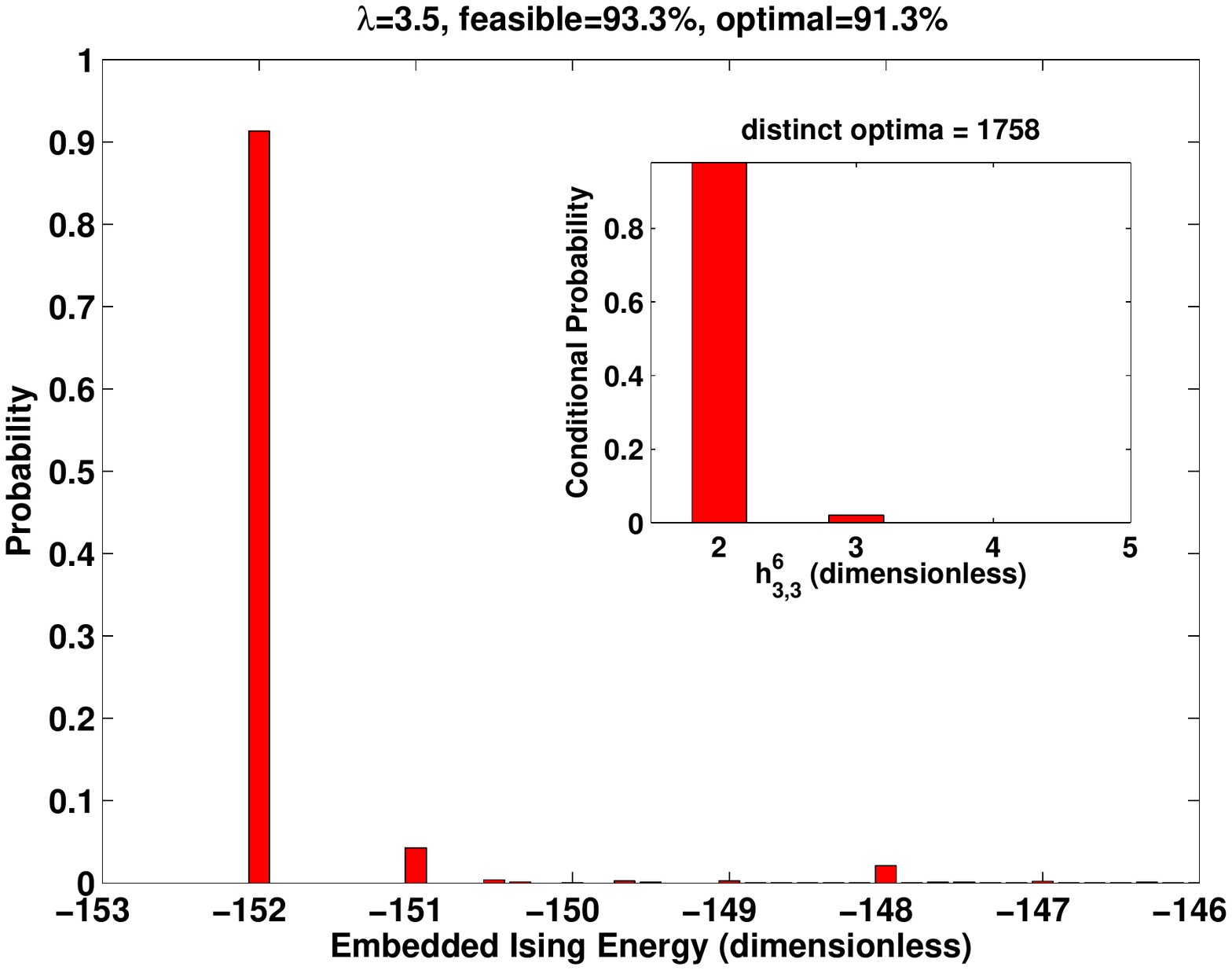} } }
\end{center}
\caption{\textbf{Energy histograms for $R(3,3)$ with $N=4,5,6$.}}
\label{R33Fig}
\end{figure}

As in the manuscript, two histograms are given for each $N$: the main
histogram corresponds to spin energies that include contributions from the
Ramsey energy function and the energy penalty functions that enforce the
embedding of the primal graph vertices into the chip. As described above, the
energy in the inset histograms is the Ramsey energy $h^{N}_{m,n}(\vc{a})$.
The energy histograms for $R(3,3)$ appear in Fig.~\ref{R33Fig}.

Examining the inset histogram for $N=4$ we see that: (i)~$h_{min}=0$;
and (ii)~$18$ distinct $\vc{a}$-configurations have zero-energy, corresponding
to $18$ distinct graphs with no $3$-cliques or $3$-independent sets. From the
main histogram we see that the probability to find an optimal
$\vc{a}$-configuration is approximately $0.93$. Note that $h_{min}=0$ is
exactly the $N=4$ final ground-state energy $E_{gs}(t_{f})=0$ found in
Ref.~\onlinecite{G&C2011}, indicating that quantum annealing (QA) finds
the final ground-state with high probability. The number of optimal graphs
found agrees with that found numerically in Ref.~\onlinecite{G&C2011}.

A similar examination of the main histogram for $N=5$ gives: (i)~$h_{min}=0$;
and (ii)~there are $12$ optimal graphs/$\vc{a}$-configurations. From the main
histogram we see that the probability to find an optimal $\vc{a}$-configuration
is approximately $0.94$. Again, $h_{min}=0$ equals the $N=5$ ground-state
energy $E_{gs}(t_{f})=0$ found in Ref.~\onlinecite{G&C2011}, and so QA finds
the final ground-state with high probability. The number of optimal graphs found
experimentally agrees with that found numerically in Ref.~\onlinecite{G&C2011}.

For $N=6$, we see that: (i)~$h_{min}=2$; and (ii)~$1758$ optimal graphs
were observed. From the main histogram we see that the probability to find an
optimal $\vc{a}$-configuration is approximately $0.91$. As shown in
Ref.~\onlinecite{G&C2011}, $E_{gs}(t_{f})=2$ for $N=6$, and so $h_{min}=
E_{gs}(t_{f})$, and QA again finds the final ground-state with high probability.
Since $N=6$ is the first $N$ value for which $h_{min}=E_{gs}(t_{f})>0$, the
protocol for the Ramsey quantum algorithm correctly\cite{Bollobas} identifies
$R(3,3)=6$. Finally, Ref.~\onlinecite{G&C2011} showed that
the number of optimal graphs for $N=6$ is $1760$ so that QA found all but two
of the $1760$ optimal graphs.

Table~\ref{table1} summarizes all of the above results.

\subsection{$\mathbf{R(m,2)=m}$}
\label{sec1b}

Here we present our experimental results for $R(m,2)$ with $4\leq m\leq 8$. 
Since the $R(m,2)$ discussion closely parallels that of $R(3,3)$, we give a 
more abbreviated presentation. In Section~\ref{energyfuncs} above we showed
that for $N < m$, the cost function $h^{N}_{m,2}
(\vc{a})$ is linear in $\vc{a}$ and consequently the spins are uncoupled when
this cost function is translated to spin variables $\vc{s}_{a}$. There is thus
no need to introduce ancillary $\vc{s}_{b}$-spins, nor do we need to include
ferromagnetic equality penalties in the Ising model. For this reason all spin
configurations are feasible for $N<m$, and no $\lambda$ values are recorded
in the main histograms. All results below are included in Table~\ref{table1}.\\

\begin{figure}
\begin{center}
\mbox{\subfigure[$\; N=3$]{\includegraphics
[trim= 0 6cm 0 7cm, clip, width=9.5cm]{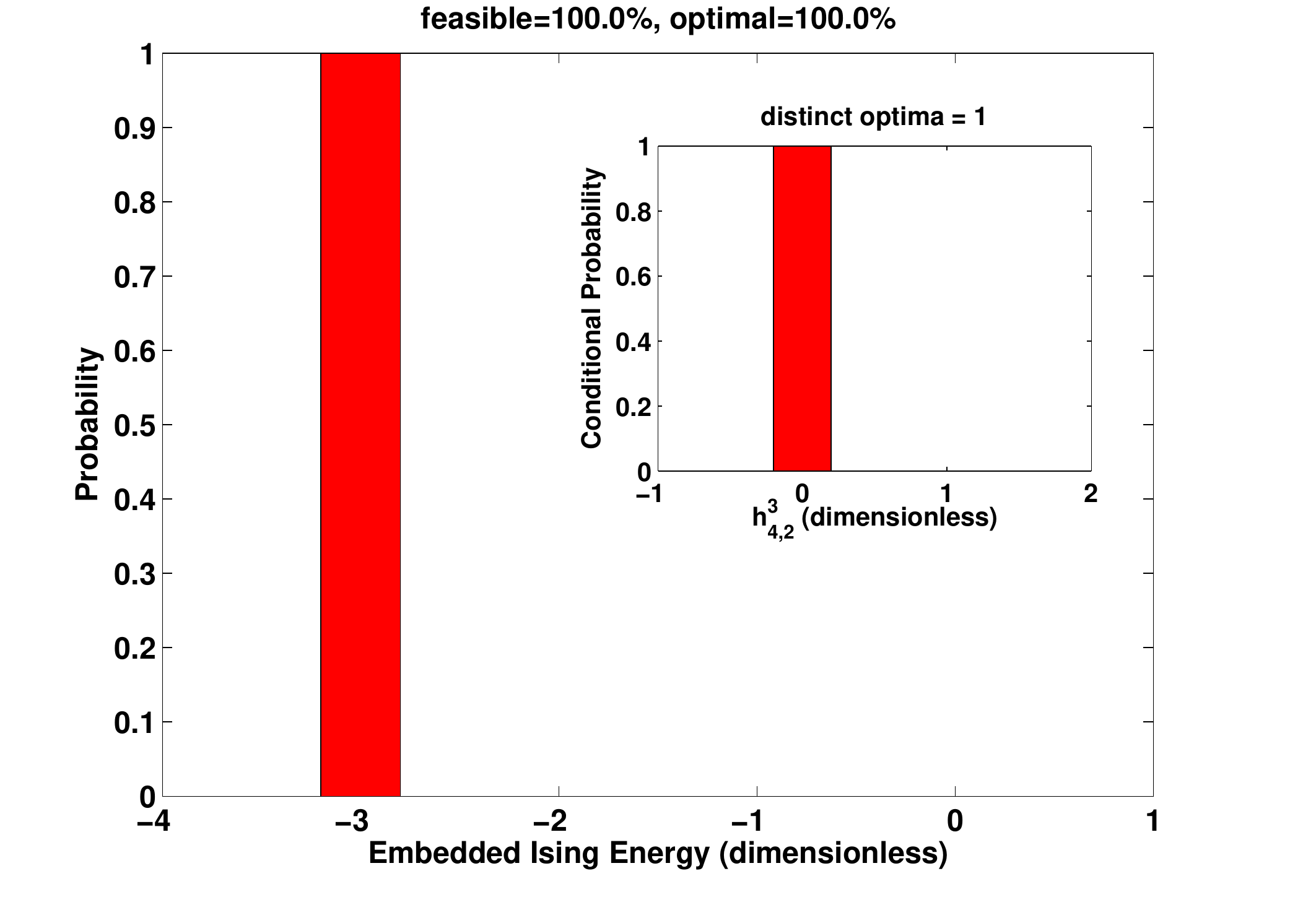}} }
\mbox{\subfigure[$\; N=4$]{\includegraphics
[trim= 0 6cm 0 6cm, clip, width=9.5cm]{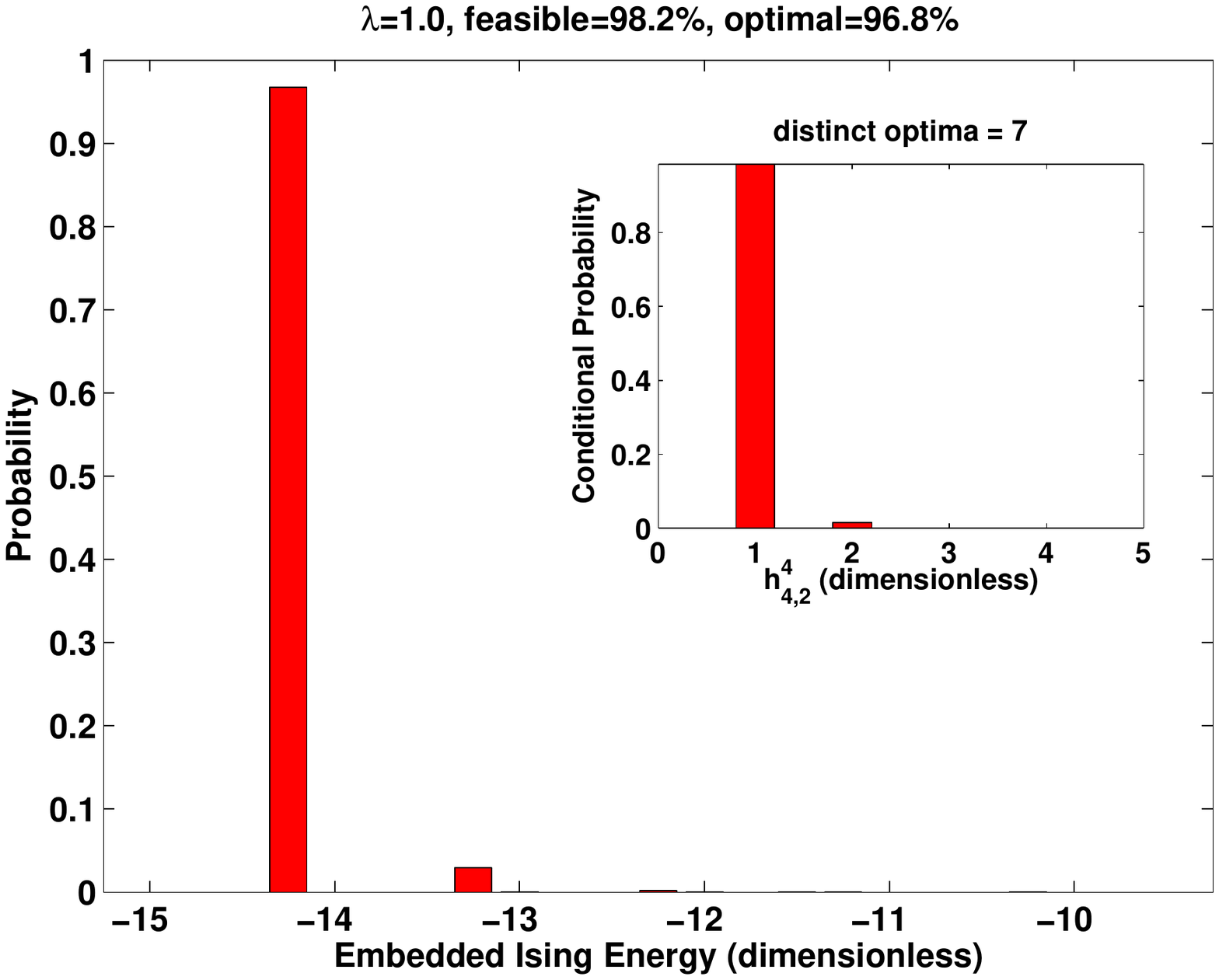}} }
\end{center}
\caption{\textbf{Energy histograms for $R(4,2)$ with $N=3$ and $4$.}}
\label{R42Fig}
\end{figure}

$\mathbf{R(4,2):}$ Fig.~\ref{R42Fig}
contains the energy histograms for $R(4,2)$. Examining the inset histogram for
$N=3$ ($4$), we see that: (i)~$h_{min}= 0$ ($1$); (ii)~the number of optimal
graphs is $1$ ($7$); and (iii)~from the main histogram,  the probability to find
an optimal $\vc{a}$-configuration is approximately $1.0$ ($0.97$). The minimum
energies $h_{min} =0$ and $1$ for $N=3$ and $4$ agree exactly with
the corresponding final ground-state energies $E_{gs}(t_{f})$ found in
Ref.~\onlinecite{G&C2011}, indicating that QA finds the final ground-state
with high probability. As $h_{min}$ jumps from $0\rightarrow 1$ as $N$ goes
from $3\rightarrow 4$, the Ramsey protocol correctly\cite{Bollobas} identifies
$R(4,2)=4$. Finally, Ref.~\onlinecite{G&C2011} showed that the number of
optimal graphs for $N=3$ and $4$ are, respectively, $1$ and $7$, which agrees
with the number of optimals found by QA. For $N=3$, the unique optimal graph
is the fully connected $3$-vertex graph which has no $4$-cliques or
$2$-independent sets, while for $N=4$, the optimal graphs are the
$\binom{4}{2}=6$ graphs with a single $2$-independent set and no
$4$-clique, and the unique fully connected $4$-vertex graph which has
one $4$-clique and no $2$-independent set.\\

\begin{figure}
\begin{center}
\mbox{\subfigure[\; $N=4$]{\includegraphics
[trim= 0 6cm 0 7cm, clip, width=9.5cm]{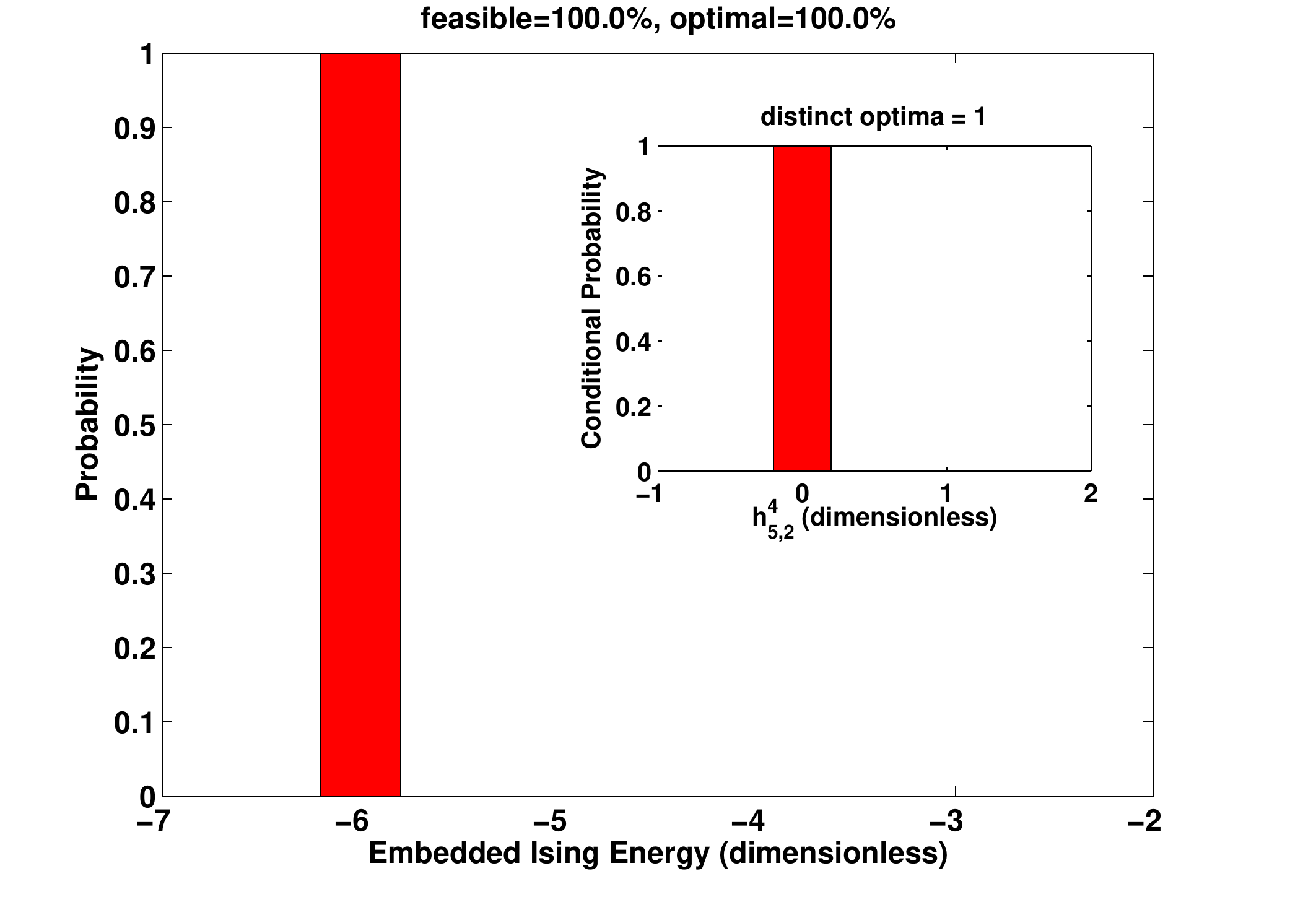}} }
\mbox{\subfigure[\; $N=5$]{\includegraphics
[trim= 0 6cm 0 6cm, clip, width=9.5cm]{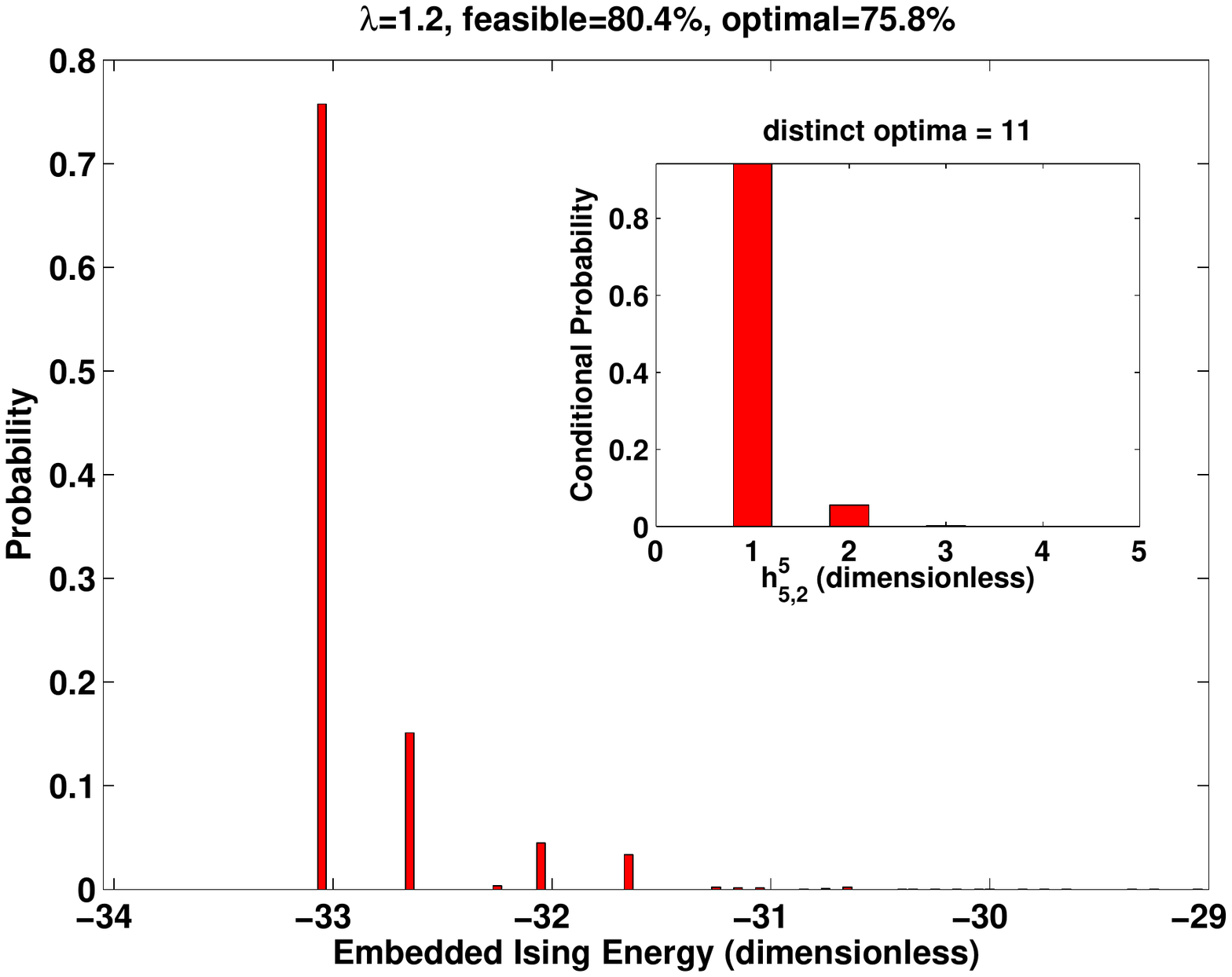}} }
\end{center}
\caption{\textbf{Energy histograms for $R(5,2)$ with $N=4$ and $5$.}}
\label{R52Fig}
\end{figure}

$\mathbf{R(5,2):}$ Fig.~\ref{R52Fig}
contains the energy histograms for $R(5,2)$. Examining the inset histogram for
$N=4$ ($5$), we see that: (i)~$h_{min}= 0$ ($1$); (ii)~the number of optimal
graphs is $1$ ($11$); and (iii)~from the main histogram,  the probability to
find an optimal $\vc{a}$-configuration is approximately $1.0$ ($0.76$). The
minimum energies $h_{min} =0$ and $1$ for $N=4$ and $5$ agree exactly with
the corresponding final ground-state energies $E_{gs}(t_{f})$ found in
Ref.~\onlinecite{G&C2011}, indicating that QA finds the final ground-state
with high probability. As $h_{min}$ jumps from $0\rightarrow 1$ as $N$ goes
from $4\rightarrow 5$, the Ramsey protocol correctly\cite{Bollobas} identifies
$R(5,2)=5$. Finally, Ref.~\onlinecite{G&C2011} showed that
the number of optimal graphs for $N=4$ and $5$ are, respectively, $1$ and
$11$, which agrees with the number of optimals found by QA. For $N=4$, the
unique optimal graph is the fully connected $4$-vertex graph which has no
$5$-cliques or $2$-independent sets, while for $N=5$, the optimal graphs are
the $\binom{5}{2}=10$ graphs with a single $2$-independent set and no
$5$-clique, and the unique fully connected $5$-vertex graph which has
one $5$-clique and no $2$-independent set.\\

\begin{figure}
\begin{center}
\mbox{\subfigure[\; $N=5$]{\includegraphics
[trim= 0 6cm 0 7cm, clip, width=9.5cm]{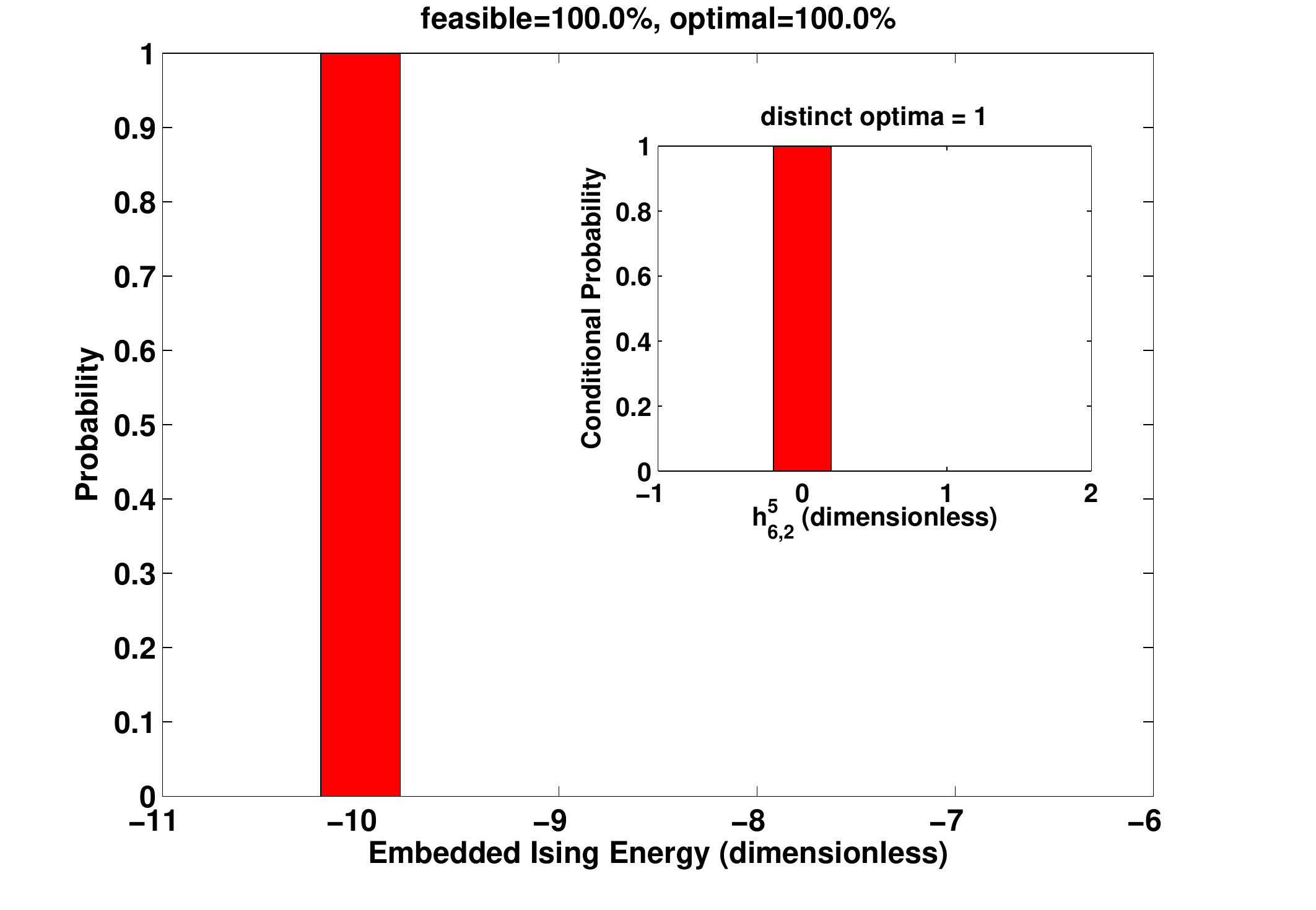}} }
\mbox{\subfigure[\; $N=6$]{\includegraphics
[trim= 0 6cm 0 6cm, clip, width=9.5cm]{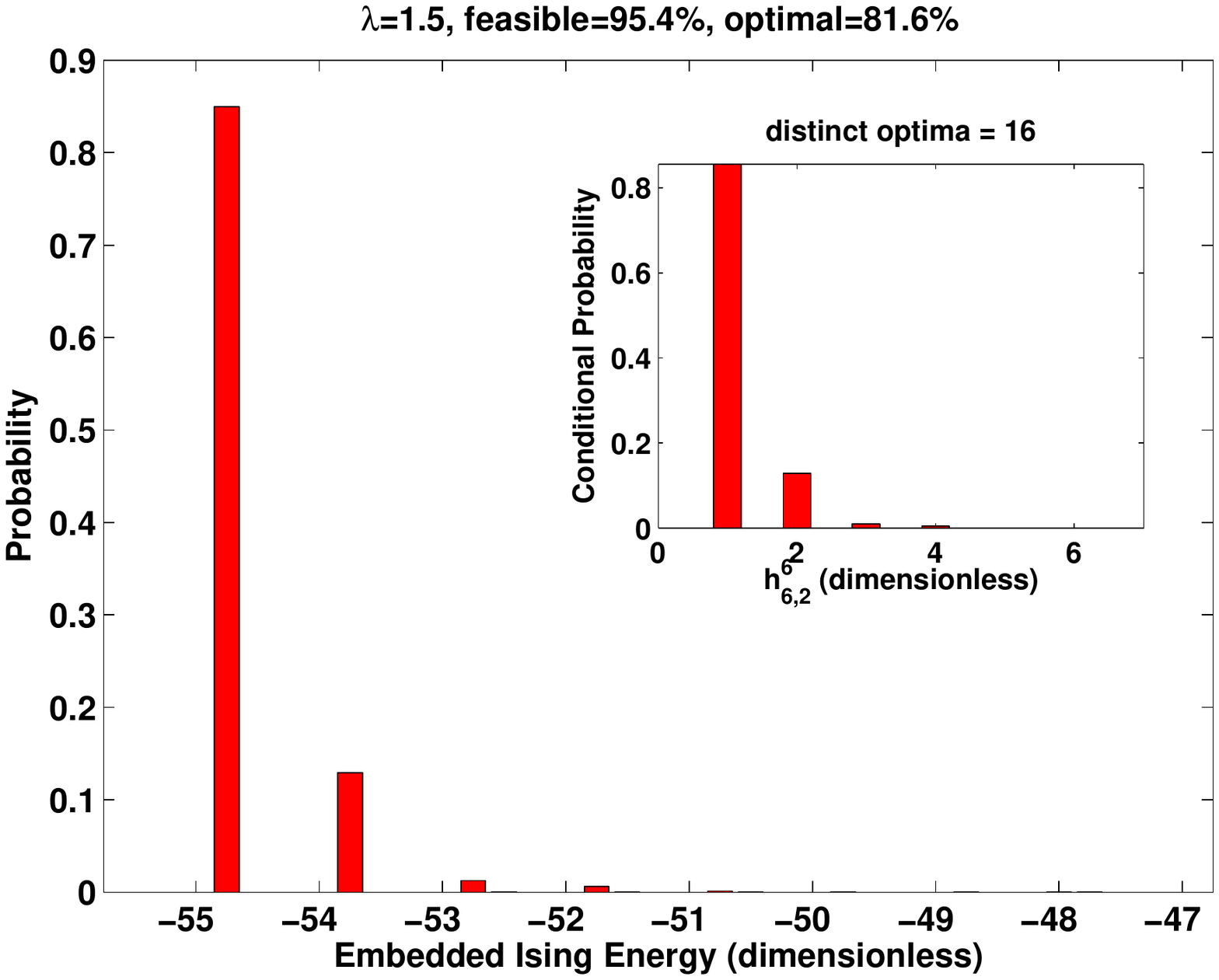}} }
\end{center}
\caption{\textbf{Energy histograms for $R(6,2)$ with $N=5$ and $6$.}}
\label{R62Fig}
\end{figure}

$\mathbf{R(6,2):}$ Fig.~\ref{R62Fig}
contains the energy histograms for $R(6,2)$. Examining the inset histogram for
$N=5$ ($6$), we see that: (i)~$h_{min}= 0$ ($1$); (ii)~the number of optimal
graphs is $1$ ($16$); and (iii)~from the main histogram, the probability to find
an optimal $\vc{a}$-configuration is approximately $1.0$ ($0.82$). The
minimum energies $h_{min} =0$ and $1$ for $N=5$ and $6$ agree exactly
with the corresponding final ground-state energies $E_{gs}(t_{f})$ found in
Ref.~\onlinecite{G&C2011}, indicating that QA finds the final ground-state
with high probability. As $h_{min}$ jumps from $0\rightarrow 1$ as $N$ goes
from $5\rightarrow 6$, the Ramsey protocol correctly\cite{Bollobas} identifies
$R(6,2)=6$. Finally, Ref.~\onlinecite{G&C2011} showed that the number of
optimal graphs for $N=5$ and $6$ are, respectively, $1$ and $16$, which agrees
with the number of optimals found by QA. For $N=5$, the unique optimal graph
is the fully connected $5$-vertex graph which has no $6$-cliques or
$2$-independent sets, while for $N=6$, the optimal graphs are the
$\binom{6}{2}=15$ graphs with a single $2$-independent set and no $6$-clique,
and the unique fully connected $6$-vertex graph which has one $6$-clique and no
$2$-independent set.\\

\begin{figure}
\begin{center}
\mbox{\subfigure[\; $N=6$]{\includegraphics
[trim= 0 6cm 0 7cm, clip, width=9.5cm]{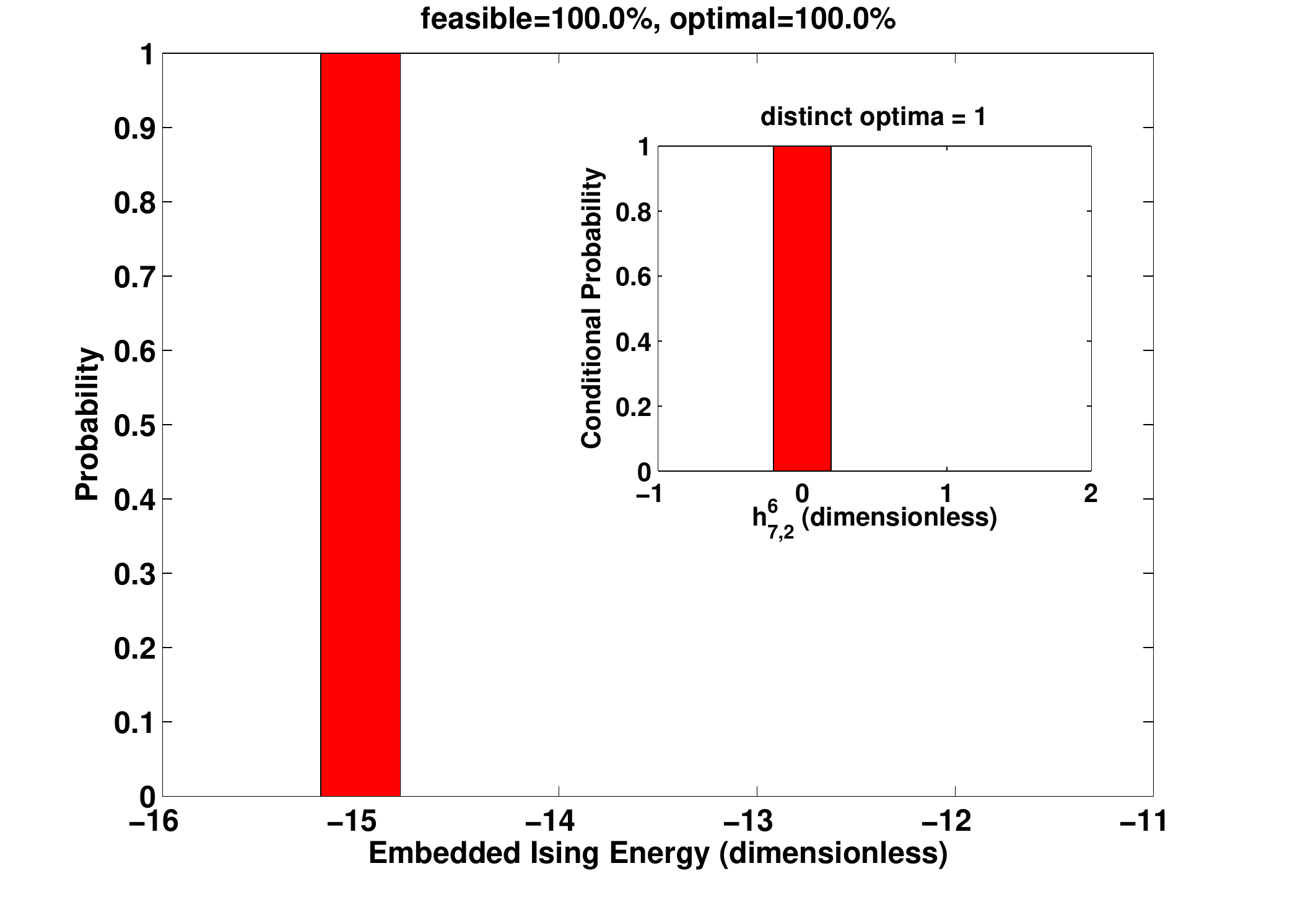}} }
\mbox{\subfigure[\; $N=7$]{\includegraphics
[trim= 0 6cm 0 6cm, clip, width=9.5cm]{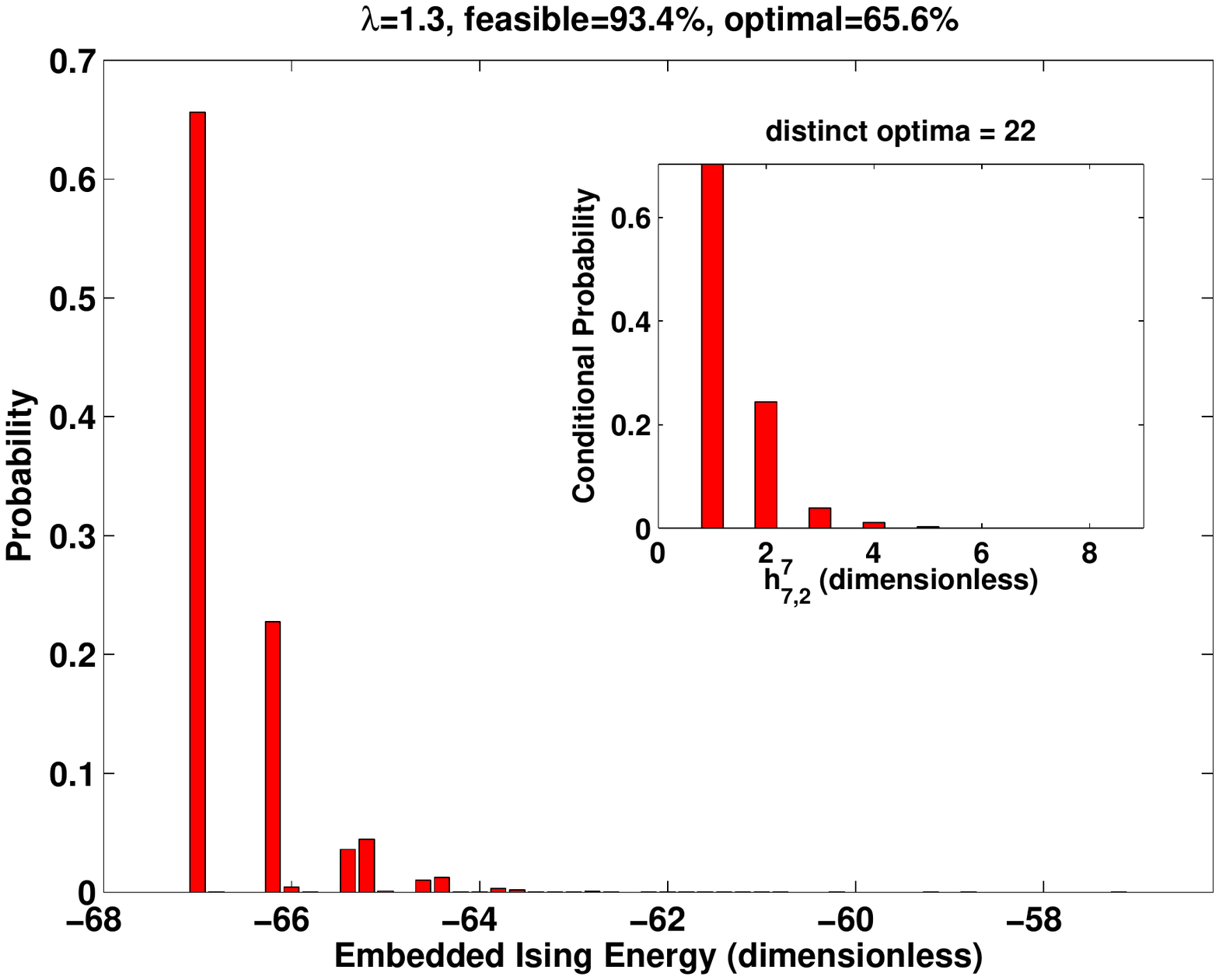}} }
\end{center}
\caption{\textbf{Energy histograms for $R(7,2)$ with $N=6$ and $7$.}}
\label{R72Fig}
\end{figure}

$\mathbf{R(7,2):}$ Fig.~\ref{R72Fig}
contains the energy histograms for $R(7,2)$. Examining the inset histogram for
$N=6$ ($7$), we see that: (i)~$h_{min}= 0$ ($1$); (ii)~the number of optimal
graphs is $1$ ($22$); and (iii)~from the main histogram, the probability to find
an optimal $\vc{a}$-configuration is approximately $1.0$ ($0.66$). The minimum
energies $h_{min} =0$ and $1$ for $N=6$ and $7$ agree with
the corresponding final ground-state energies $E_{gs}(t_{f})$ found in
Ref.~\onlinecite{G&C2011}, indicating that QA finds the final ground-state with
high probability. As $h_{min}$ jumps from $0\rightarrow 1$ as $N$ goes from
$6\rightarrow 7$, the Ramsey protocol correctly\cite{Bollobas} identifies
$R(7,2)=7$. Finally, Ref.~\onlinecite{G&C2011} showed that the number of
optimal graphs for $N=6$ and $7$ are, respectively, $1$ and $22$, which agrees
with the number of optimals found by QA. For $N=6$, the unique optimal graph
is the fully connected $6$-vertex graph which has no $7$-cliques or
$2$-independent sets, while for $N=7$, the optimal graphs are the
$\binom{7}{2}=21$ graphs with a single $2$-independent set and no
$7$-clique, and the unique fully connected $7$-vertex graph which has
one $7$-clique and no $2$-independent set.\\

\begin{figure}
\begin{center}
\mbox{\subfigure[$\; N=6$]{\includegraphics
[trim= 0 6cm 0 7cm, clip, width=9.5cm]{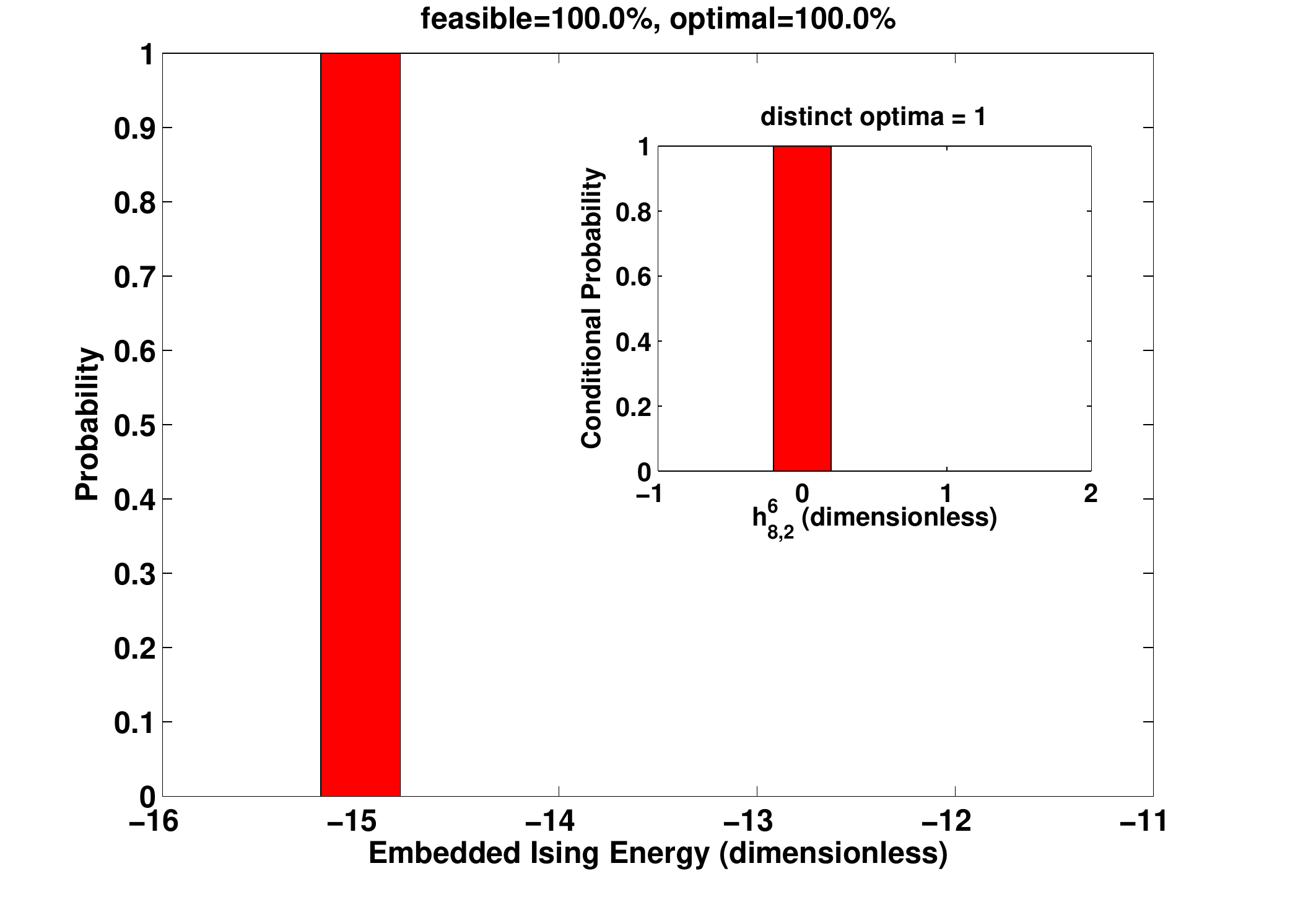} } }
\mbox{\subfigure[$\; N=7$]{\includegraphics
[trim= 0 6cm 0 6cm, clip, width=9.5cm]{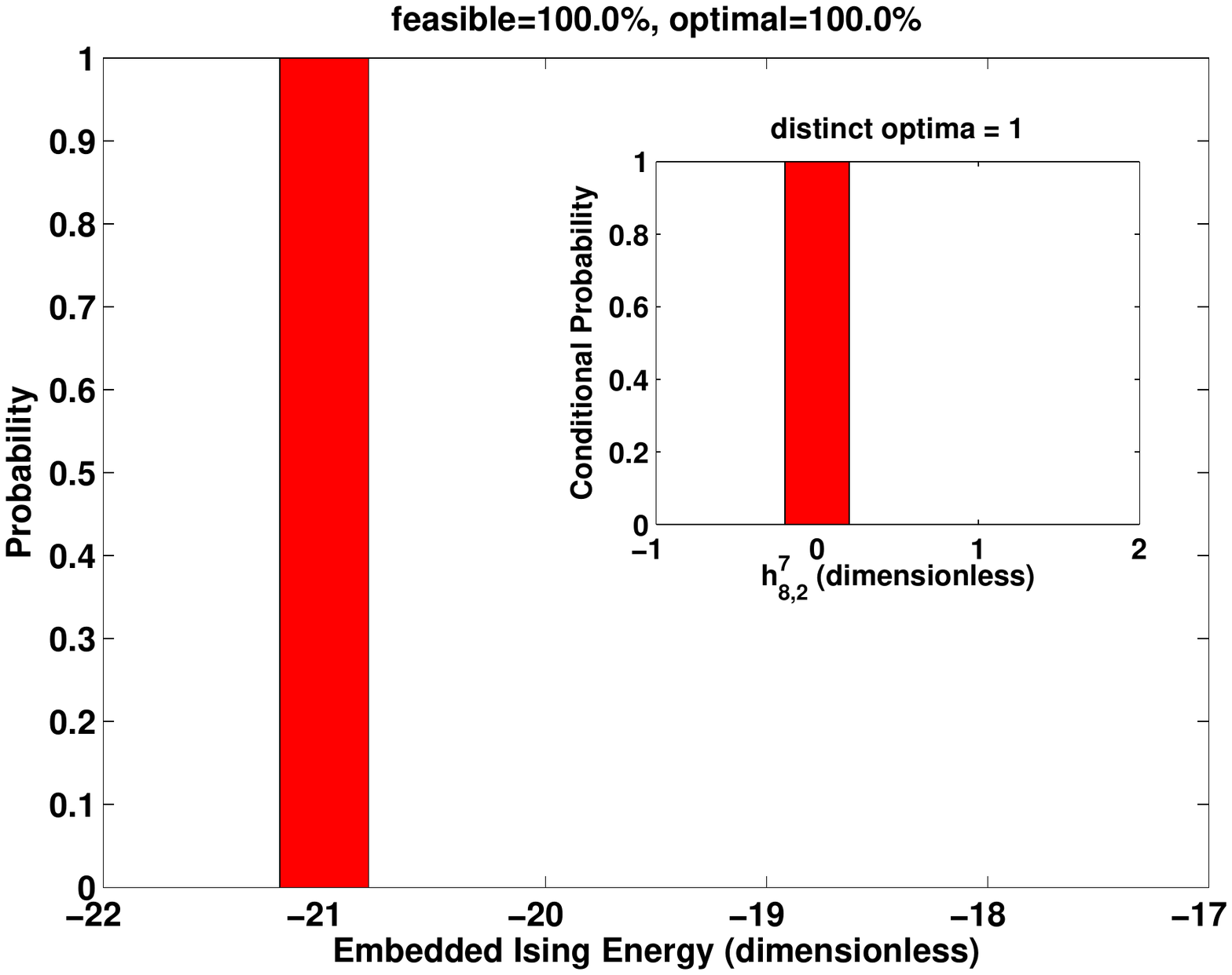} } }
\mbox{\subfigure[$\; N=8$]{\includegraphics
[trim= 0 6cm 0 6cm, clip, width=9.5cm]{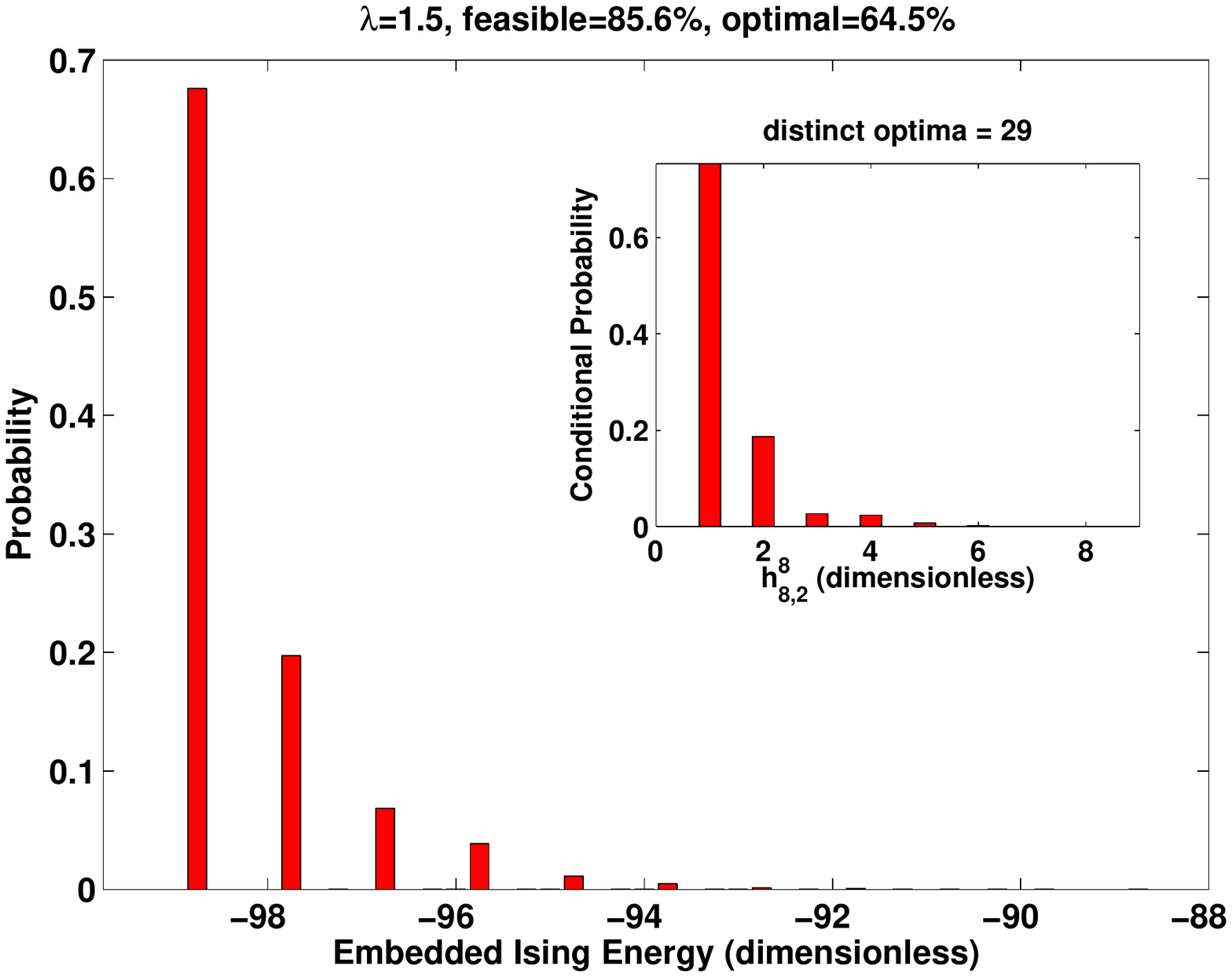} } }
\end{center}
\caption{\textbf{Energy histograms for $R(8,2)$ with $N=6,7,8$.}}
\label{R82Fig}
\end{figure}

$\mathbf{R(8,2):}$ Fig.~\ref{R82Fig}
contains the energy histograms for $R(8,2)$. 
Examining the inset histogram for $N=6$ we see that: (i)~$h_{min}=0$;
and (ii)~a single $\vc{a}$-configuration has zero-energy, corresponding
to the fully connected $6$-vertex graph which has no $8$-clique or 
$2$-independent sets. From the main histogram we see that the probability 
to find an optimal $\vc{a}$-configuration is $1.0$. Note that $h_{min}=0$ is
exactly the $N=6$ final ground-state energy $E_{gs}(t_{f})=0$ found in
Ref.~\onlinecite{G&C2011}, indicating that quantum annealing (QA) finds
the final ground-state with high probability. The number of optimal graphs
found experimentally agrees with that found in Ref.~\onlinecite{G&C2011}.

A similar examination of the main histogram for $N=7$ gives: (i)~$h_{min}=0$;
and (ii)~a single optimal graph/$\vc{a}$-configuration. The unique optimal 
graph is the fully connected $7$-vertex graph which has no $8$-clique or 
$2$-independent set. From the main histogram we see that the probability 
to find an optimal $\vc{a}$-configuration is $1.0$. Again, $h_{min}=0$ equals 
the $N=7$ ground-state energy $E_{gs}(t_{f})=0$ found in 
Ref.~\onlinecite{G&C2011}, and so QA finds the final ground-state with high 
probability.  The number of optimal graphs found experimentally agrees exactly 
with that found in Ref.~\onlinecite{G&C2011}.

For $N=8$, we see that: (i)~$h_{min}=1$; and (ii)~$29$ optimal graphs
were observed. Here the optimal graphs are the $\binom{8}{2}=28$ graphs 
with a single $2$-independent set and no $8$-clique, and the unique fully 
connected $8$-vertex graph which has one $8$-clique and no $2$-independent 
set. From the main histogram we see that the probability to find an
optimal $\vc{a}$-configuration is approximately $0.65$. As shown in
Ref.~\onlinecite{G&C2011}, $E_{gs}(t_{f})=1$ for $N=8$, and so $h_{min}=
E_{gs}(t_{f})$, and QA again finds the final ground-state with high probability.
Since $N=8$ is the first $N$ value for which $h_{min}=E_{gs}(t_{f})>0$, the
protocol for the Ramsey quantum algorithm correctly\cite{Bollobas} identifies
$R(8,2)=8$. Finally, Ref.~\onlinecite{G&C2011} showed that the number of 
optimal graphs for $N=8$ is $29$ so that QA found all of the optimal graphs.\\

\section{Embedding primal graph into chip hardware}
\label{sec2}

Fig.~\ref{qubitFig} reproduces Fig.~1 of the manuscript which is included here 
for convenience. As discussed in Section~\ref{impcost}, the primal graph of an 
optimization cost function consists of: (i)~vertices that represent bit variables 
whose values are to be optimized; and (ii)~edges that connect pairs of interacting 
bits. The primal graph of $h^N_{m,n}(\vc{s})$ is the same as the primal graph 
of $h^N_{m,n}(\vc{a})$ as the former is obtained from the latter by the 
substitution $\vc{s}=2\vc{a}-1$.

\subsection{R(8,2)}
\label{sec2b}

\begin{figure}
\begin{center}
\tikzstyle {var} = [circle,draw=blue!50,fill=blue!10,thick,minimum size=1.1cm]
\begin{tikzpicture}[>=latex,text height=1.5ex,text depth=0.25ex]
\matrix[row sep=0.6cm,column sep=0.9cm] {
&
\node (b2) [var] {$b_2$}; &
\node (b3) [var] {$b_3$}; &
\node (uDots)  {$\cdots$}; &
\node (bL1) [var] {$b_{L-1}$}; &
&
\\
\node (a1) [var] {$a_1$}; &
\node (a2) [var] {$a_2$}; &
\node (a3) [var] {$a_3$}; &
\node (lDots)  {$\cdots$}; &
\node (aL1) [var] {$a_{L-1}$}; &
\node (aL) [var] {$a_L$};
\\
};
\draw (a1) edge (b2);
\draw (a2) edge (b2);
\draw (b2) edge (b3);
\draw (a2) edge (b3);
\draw (b3) edge (uDots);
\draw (a3) edge (b3);
\draw (a3) edge (lDots);
\draw (uDots) edge (bL1);
\draw (lDots) edge (bL1);
\draw (aL1) edge (bL1);
\draw (bL1) edge (aL);
\draw (aL1) edge (aL);
\end{tikzpicture}
\end{center}
\caption{\textbf{Primal graph for $h^m_{m,2}(\vc{a},\vc{b})$.} $L=\binom{m}{2}$
is the total number of $\vc{a}$ variables.} \label{rm2PrimalFig}
\end{figure}
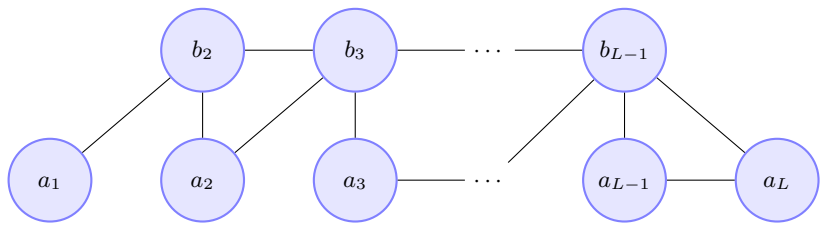

\begin{figure}
\begin{center}
\includegraphics[width=9cm]{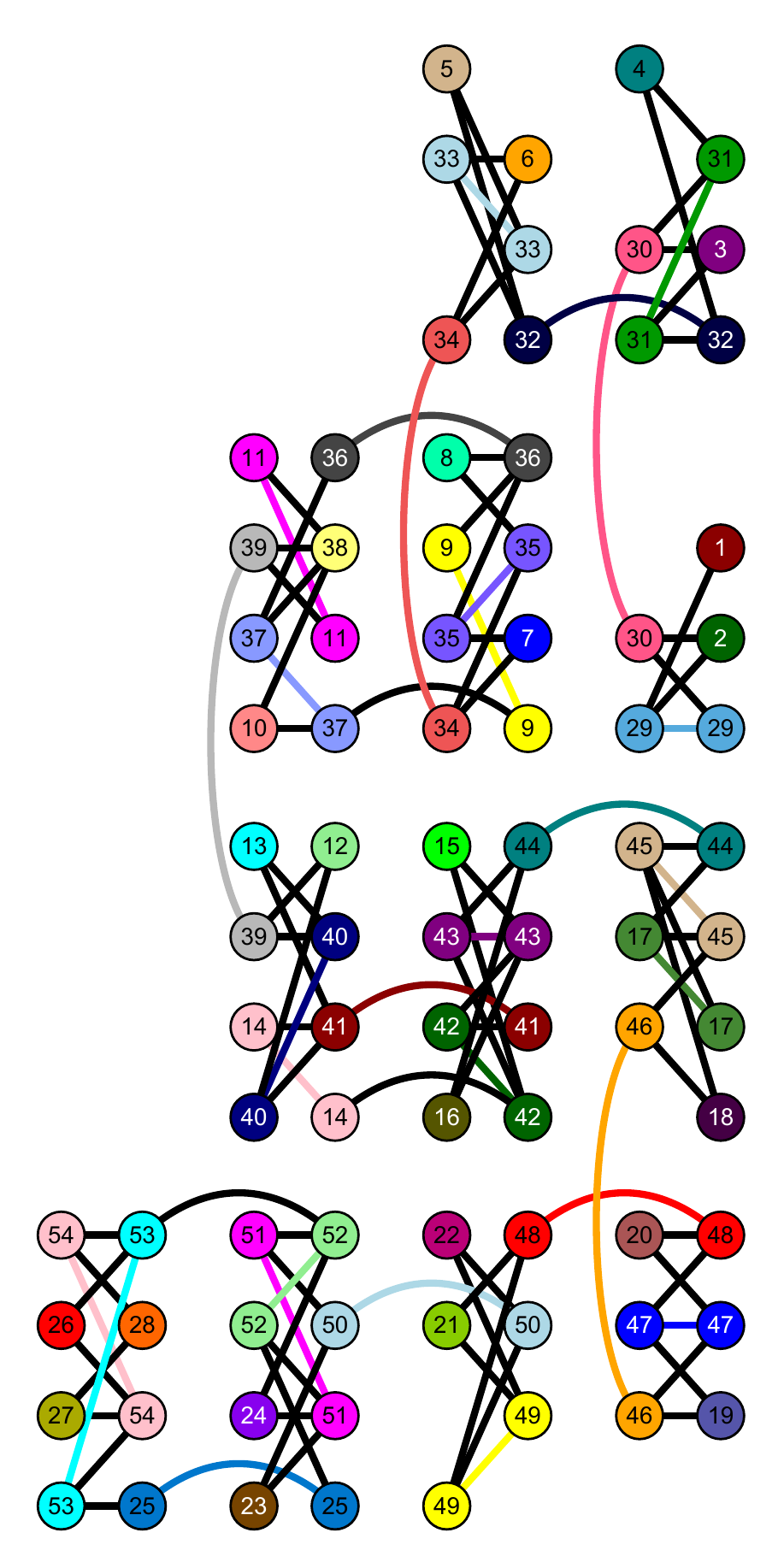}
\end{center}
\caption{\textbf{Embedding of the primal graph of Fig. \ref{rm2PrimalFig} into
hardware.}  Qubits are numbered so that qubits labeled $i$ with $1\leq i \leq 28$
correspond to primal variables $a_i$, and qubits labeled $29, \cdots , 54$
correspond to the ancillary variables $b_2, \cdots , b_{27}$.}
\label{r82EmbedFig}
\end{figure}

The Ramsey cost function $h^8_{8,2}(\vc{a})$ is quite complicated as it involves
a term that couples 28 spin variables. As discussed in Section~\ref{energyfuncs}
above this requires the introduction of 26 ancillary $\vc{b}$-variables to 
reduce this product to a sum of pairwise interactions. The resulting primal 
graph for $h^8_{8,2}(\vc{a},\vc{b})$ is shown in Fig.~\ref{rm2PrimalFig}. For 
this case there are a total of 54 primal $\vc{a}$ and $\vc{b}$ variables.
The embedding of the primal graph in Fig.~\ref{rm2PrimalFig} into the chip is
shown in Fig.~\ref{r82EmbedFig}. Note that an additional $30$ qubits are
needed to complete the embedding, bringing the total number of qubits used
in the computation to 84. Like-colored qubits represent a single primal
variable, though note that certain colors had to be reused. As before black
edges carry the primal variable couplings. The topmost qubit labeled 5 in
Fig.~\ref{r82EmbedFig} corresponds to qubit 17 of Fig.~\ref{qubitFig}.

\subsection{$\mathbf{R(m,2)}$ for $\mathbf{4\leq m\leq 7}$}
\label{sec2c}

The procedure for finding the chip embeddings for the Ramsey cost functions
$h^m_{m,2}$ with $m = 4, \cdots , 7$ is similar to that presented for
$h^{8}_{8,2}$. The resulting embeddings are much simpler and are not
reproduced here.

\subsection{R(3,3)}
\label{sec2a}
\begin{figure}
\begin{center}
\includegraphics[width=7cm]{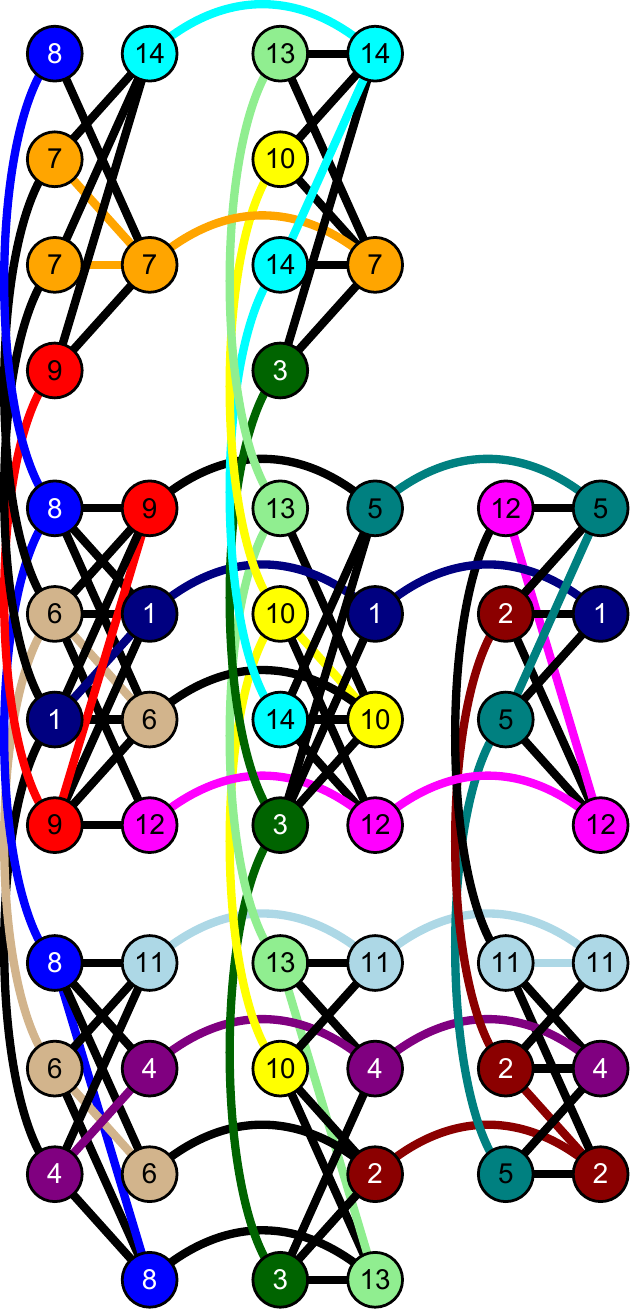}
\end{center}
\caption{\textbf{Embedding of the primal graph of $h^6_{3,3}(0,a_2,a_3,\cdots,
a_{15})$ into hardware.} Variable $a_{i+1}$ is labeled as qubit $i$.}
\label{r33EmbedFig}
\end{figure}

As explained in Section~$3$ above, the cost function $h^6_{3,3}
(\vc{a})$ has 14 variables since we fixed $a_1 =0$.  Fig.~\ref{r33EmbedFig}
shows the embedding of the primal graph for the problem $h^6_{3,3}(0,a_2,a_3,
\cdots,a_{15})$. Note that the qubit labeled $i$ in this Figure corresponds to the
binary variable $a_{i+1}$. Like-colored qubits are connected with ferromagnetic
couplings of strength $\lambda$ along the like-colored edges. Black edges are
used to represent the coupling strengths between primal variables.The embedding
is situated within the chip so that the top-left qubit labeled as 8 in
Fig.~\ref{r33EmbedFig} corresponds to qubit 41 of Fig.~\ref{qubitFig}.
Note that many other embeddings into the chip are possible, and we make no
claims that this embedding uses the smallest numbers of qubits to represent the
required connectivity.

\section{Discussion}
\label{sec4}
In this Section we discuss a number of important topics related to the Ramsey 
number experiments and their analysis.

\subsection{Parameter noise}
\label{sec4a}

Parameter noise enters during the programming of the quantum annealing (QA) 
Hamiltonian $H(t)$ onto the chip and this noise is frozen into the actual value 
the local fields $\bldh$ and coupling constants $\bldJ$ take on. By this we 
mean that the actual value of (say) the local field $\bldh_{a} = \bldh_{0} +
\Delta\bldh$, where $\bldh_{0}$ is the nominal value we would like $\bldh$ to 
take plus an error $\Delta\bldh$. It is important to understand that, because 
the chip is programmed only once, $\Delta\bldh$ does not change from one QA 
run to the next. Thus the effect of parameter noise is to produce a small static
random shift in $\bldh$ and $\bldJ$ away from nominal values $\bldh_{0}$ and 
$\bldJ_{0}$, but which remains fixed from one QA run to the next. Now if the 
lowest-lying instantaneous energies $E(t)$ associated with the QA Hamiltonian 
$H(t)$ vary rapidly with small changes in $\bldh$ and $\bldJ$, then hardware 
performance on the optimization problem associated with $H(t)$ can be expected 
to be sensitive to parameter noise. The indication of this would be disagreement 
between the theoretical predictions based on $\bldh = \bldh_{0}$ and $\bldJ = 
\bldJ_{0}$, and the experimental results based on the shifted values 
$\bldh_{a}$ and $\bldJ_{a}$. On the other hand, if the low-lying energy landscape 
does not vary rapidly with small changes in $\bldh$ and $\bldJ$, then hardware 
performance should be well-described by the nominal parameters $\bldh_{0}$ and 
$\bldJ_{0}$ and so theoretical predictions and experimental results should have 
strong agreement, and hardware performance should not be sensitive to parameter 
noise.

Note that in the Ramsey number experiments reported in the mansucript and in 
this SI: (i)~all Ramsey numbers $R(m,2)$ with $4\leq m \leq 8$ and $R(3,3)$ were
found correctly; (ii)~all final ground-state energies $E_{gs}(T)$ were found 
correctly; and (iii)~for $R(m,2)$, and for $R(3,3)$ with $N=4$ and $5$, all 
Ramsey ground-states  were found correctly, and for $R(3,3)$ with $N=6$, 
$1758$ out of $1760$ ground-states were found correctly. It is possible to 
conclude this as exact theoretical results were available to check against 
experimental results \cite{G&C2011}. As noted above, the excellent agreement 
between theory and experiment is an indication that the Ramsey number 
experiments are not very sensitive to parameter noise. Another way to 
understand this lack of sensitivity is to note that the Ramsey energy functions 
appearing in Section~\ref{energyfuncs} above all involve $\bldh$ and $\bldJ$ 
values that are small integers, and these small integer values are safely within
the hardware's available parameter precision. 

\subsection{Distribution of final energies}
\label{sec4b}

\begin{figure}
\begin{center}
\includegraphics
[trim= 0 6cm 0 6cm, clip, width=12cm]{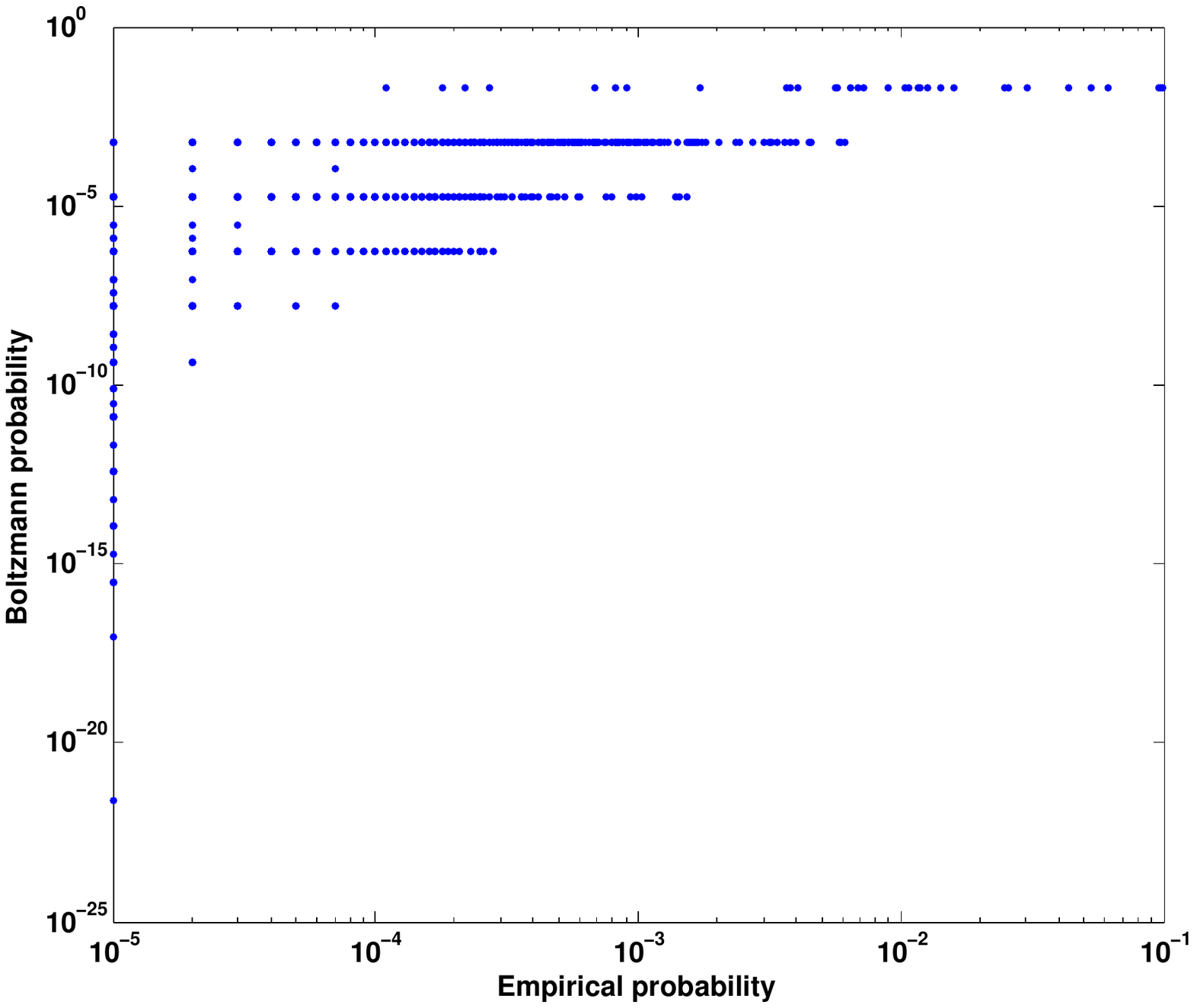}
\end{center}
\caption{Plot of Boltzmann probability versus empirical probability for $R(8,2)$
with $N=8$. See text for full discussion.}
\label{paraplot}
\end{figure}

Recall from Table I above that the final Ramsey 
ground-state (GS) is highly degenerate so that the final gap vanishes: 
$\Delta (t_{f}) = 0$. Thus as $t \rightarrow t_{f}$, the lowest energy-levels get 
arbitrarily close to each other and so transitions out of the GS become more 
and more likely. However, as $t \rightarrow t_{f}$, there remains little time for 
the hardware to relax back to the final GS before the quantum annealing (QA) 
run completes. Thus one can anticipate that the final distribution of energies 
will not correspond to a thermal equilibrium distribution described by an 
effective temperature $T_{e}$. This is in fact the case as can be seen in 
Fig.~\ref{paraplot} 
which is a parametric plot of Boltzmann versus empirical probabilities for the 
$R(8,2)$ data with $N=8$. Here each data point corresponds to a particular 
spin/ancilla configuration $(\blda ,\bldb )$ whose: (i)~x-coordinate is its 
empirical probability $P_{e}$ which is the fraction of QA runs in which 
$(\blda ,\bldb )$ was the outcome of the final measurement; and 
(ii)~y-coordinate is the Boltzmann probability $P_{B}(\blda ,\bldb ) = 
\exp[-E(\blda ,\bldb )/T_{e}]/Z$. In the Boltzmann distribution, $E(\blda ,
\bldb )$ is the hardware embedded Ramsey energy including ancilla contributions;
$T_{e}$ is an effective temperature found by doing a maximum likelihood 
Boltzmann fit to the empirical data; and $Z$ is the partition function. The fit 
yielded $T_{e} = 0.2831$ in the dimensionless units of the problem Hamiltonian 
whose energy scale is set by the maximum value of the coupling constants 
$\bldJ$ which is $1$. Notice that the data points line up 
along a number of horizontal lines, with each line having the same Boltzmann 
probability $P_{B}(E)$. Each horizontal line thus corresponds to spin/ancilla 
configurations with the same Ramsey energy $E$. If the final distribution of 
energies corresponded to thermal equilibrium, spin configurations with the 
same energy $E$ would have the same empirical probability $P_{e}$, and so 
each horizontal line would collapse to a single point on the $P_{B}(E) = 
P_{e}(E)$ line. Said another way, we see that spin/ancilla configurations with 
the same energy have different empirical probabilities which indicates 
unequivocally that the final distribution of energies does not correspond to 
thermal equilibrium. As explained above, this is not unexpected given that 
the gap $\Delta(t \rightarrow t_{f}) \rightarrow 0$ so that transitions out of the 
GS become highly likely as $t \rightarrow t_{f}$, and little time is left for the 
hardware to relax back to the GS. 

\subsection{Quantum annealing rate}
\label{sec4c}

For a given Ramsey number experiment, the initial and problem Hamiltonians 
are programmed onto the chip and remain fixed throughout all $100,000$ 
quantum annealing (QA) runs. As noted in Section~2 of the manuscript, the 
time rate of change of the instantaneous Hamiltonian 
$H(t)$ is controlled through the annealing profiles $A(t)$ and $B(t)$ that drive
the QA run. Slowing the variation of $A(t)$ and $B(t)$ would make the QA run 
more adiabatic, and thus would reduce the probability of finding the hardware 
in an excited state in the final measurement, and the distribution of final 
energies closer to a thermal equilibrium distribution. The rates chosen for the 
experiments reported in the manuscript and in the SI balance the need to 
execute a QA run in a reasonable amount of time, and the need to have a 
significant fraction of the QA runs find the hardware in the final ground-state 
(GS). Inspection of the histograms in the manuscript and the SI shows that the 
probability $P_{success}$ of finding the hardware in the final GS  satisfies 
$0.645 \leq P_{success} \leq 1.000$. These probabilities are quite high and 
indicate that the choices made for $A(t)$ and $B(t)$ are a reasonable 
compromise between competing interests. Further reduction of these rates 
would act to increase the lower bound on $P_{success}$, which is already quite 
high. Note that the Ramsey experiments were carried out with an annealing 
time of $1000\mu s$. The D-Wave hardware allows annealing times as short as 
$5\mu s$, and annealing at these faster rates is found to reduce the success 
probability $P_{success}$ as expected.

\subsection{Choosing the spin-chain ferromagnetic coupling}
\label{sec4d}

For a particular Ramsey number, the experimental procedure begins by setting 
the ferromagnetic (FM) coupling parameter $\lambda$ to a small initial value. 
Then $100,000$ QA runs are done with this $\lambda$ value and the qubits 
measured at the end of each run. The fraction of these runs $F$ in which the 
equality constraints are satisfied is then determined. If $F$ is less than 
$0.85$, the value of $\lambda$ is increased and another $100,000$ QA runs 
are carried out and the qubits measured. This process continues until $F$ is 
first found to be greater than $0.85$. At this point $\lambda$ is increased one 
last time and it is this value of $\lambda$ that is used in the experiments and 
listed at the top of each histogram. As noted above, the (at most) $15\%$ 
of QA runs which did not satisfy the equality constraints are discarded as, 
for these runs, the FM chains do not act as a single ``logical'' qubit and so 
do not properly capture the physics that couples distant qubits in the manner 
required by the Ramsey problem Hamiltonian. Only the (at least) $85\%$ of 
runs which did satisfy the equality constraints properly represent the Ramsey 
Hamiltonian and so only those runs are kept as data and analyzed. That is why 
these runs were referred to as feasible spin-configurations in the manuscript 
and SI. Of these feasible spin configurations, some fraction has the lowest 
measured energy and these are the candidate ground-states (GS). The fraction 
of the $100,000$ QA runs that is feasible and lowest energy (viz.~optimal) is 
the probability listed as ``optimal = $0.\mathrm{xxx}$'' in Figures 2 and 3 of 
the paper, and Figures 3-8 of the SI. We will describe the calculation of this 
optimal probability in detail in Section~\ref{sec4e}. For the Ramsey 
numbers considered in the manuscript, we know the final GS energy in all cases 
from either numerical simulations or analytical calculations (see Ref. 3 of the
manuscript). Thus we can conclude that the experiments found the GS energy 
correctly and with high probability. In summary, the (at most) $15\%$ of QA 
runs that violated the equality constraints never influence the data that is 
analyzed to determine the experimental results reported in the manuscript 
and the SI and so cannot cause an error in any of these results.

\subsection{Probability for optimal spin-configuration}
\label{sec4e}

All histograms appearing in the manuscript and SI report the probability
that a Ramsey experiment yields an optimal Ising spin configuration: 
``$\mathrm{optimal} = 0.\mathrm{xxx}$''. For example, in Figure 2(b) of the 
manuscript, the $64.5\%$ is found as follows. We see at the top of that figure 
that of the $100,000$ QA runs, $85.6\%$ were feasible (as defined in 
Sec.~\ref{sec4d}) spin-configurations. From the inset histogram we can read off
that approximately $75\%$ of the feasible spin-configurations were also optimal
(viz.\ had the smallest energy). Thus the fraction of total runs that were 
feasible and optimal is approximately $0.856\times 0.75 = 0.642$. The actual 
value is $0.645$ which is the $64.5\%$ quoted as optimal at the top of 
Figure~2(b). This same analysis is done for each Ramsey number to calculate 
the percentage of QA runs that yield optimal and feasible spin-configurations, 
and that percentage appears at the top of the corresponding histogram.

\subsection{Classical annealing}
\label{sec4f}

\begin{figure}
\begin{center}
\includegraphics
[trim= 0 6cm 0 7cm, clip, width=14cm]{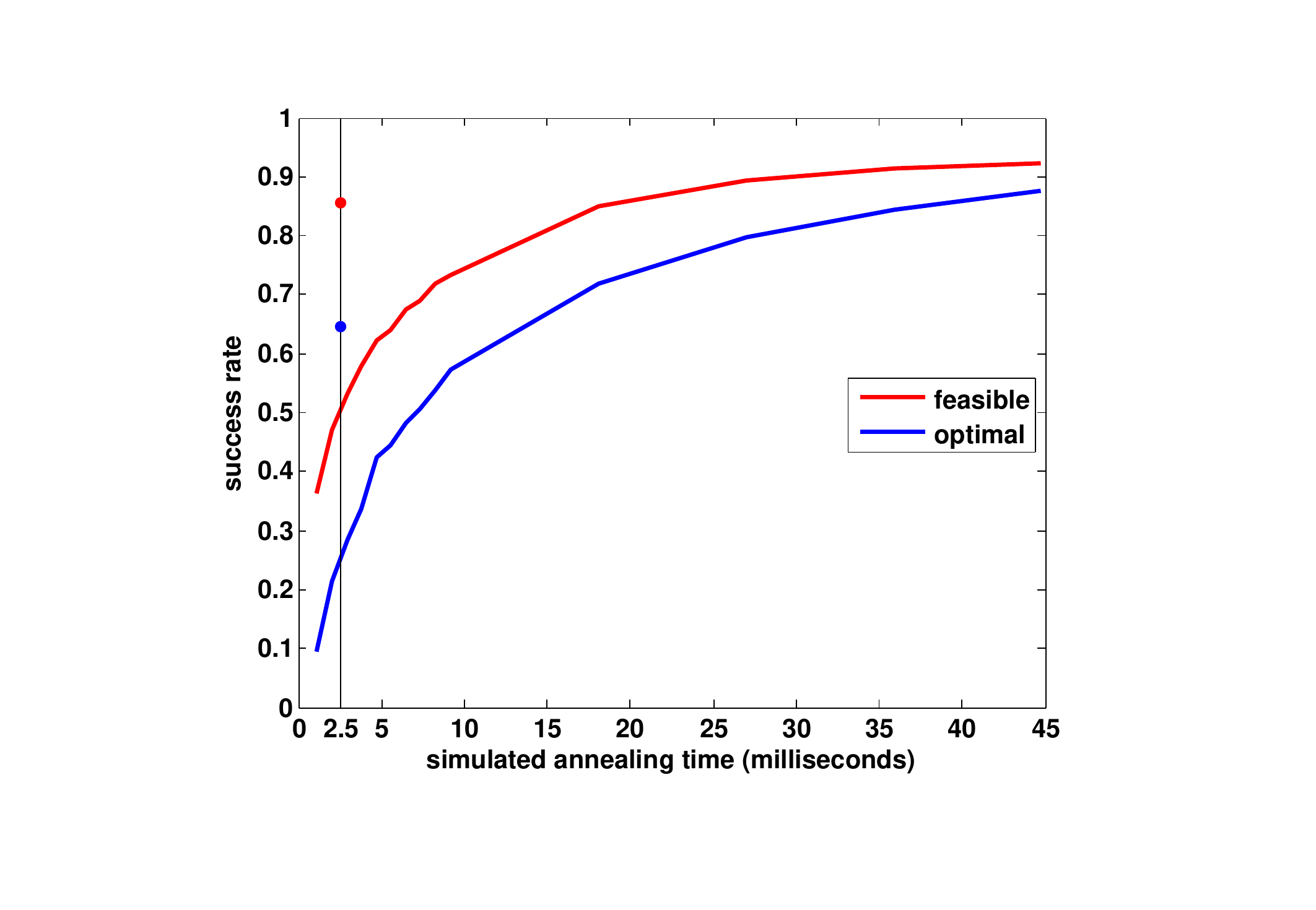}
\end{center}
\caption{Plot of simulated annealing (SA) success probability to determine 
optimal and feasible spin-configurations versus run-time for $R(8,2)$ at 
$N=8$. The SA cooling schedule is exponential; and the initial and final 
temperatures were optimized for maximum success probability. For comparison, 
quantum annealing hardware results are also shown for a runtime of $2.5ms$.
See text for further discussion.}
\label{saplot}
\end{figure}

Here we show that classical/thermal annealing can be ruled out as the source of 
optimization efficacy. First, it is clear that the hardware is not realizing classical 
annealing since the final distribution of low energy states is not Boltzmann 
distributed as discussed in Sec.~\ref{sec4b} above, and furthermore, the 
temperature 
of the hardware is never varied during the experiments. Finally, we compare the 
optimization efficacy of the quantum annealing hardware with that of an 
efficient C-implementation of simulated annealing that was run on a standard 
$8$Gb, $2.66$GHz desktop computer. The results of Fig.~\ref{saplot} 
show that at a runtime of $2.5ms$ (which is the $1ms$ runtime plus $1.5ms$ 
read-out time of the Ramsey number experiments), the quantum annealing 
hardware obtains significantly higher success rates for finding both feasible 
and optimal final spin-configurations than does simulated annealing.

\subsection{Hardware quantum coherence}
\label{sec4g}

Since 
an adiabatic quantum optimization (AQO) algorithm encodes the problem solution 
in the ground-state (GS) of the problem Hamiltonian, this class of quantum 
algorithms does not make use of a coherent superposition of instantaneous 
energy-levels. Said another way, the relative phases of the different 
instantaneous energy-levels in such a superposition contain no useful 
information about the optimization problem solution. Instead, the ability of 
the adiabatic quantum dynamics to drive the system state to the GS of the 
problem Hamiltonian directly impacts the performance of the AQO algorithm. 
Thus a more useful performance metric for an AQO algorithm is the Uhlmann 
fidelity $F = (1/\sqrt{P_{0}})\mathrm{Tr}\sqrt{\sqrt{\rho_{0}}\,\rho_{s}
\sqrt{\rho_{0}}} $ which is the overlap of the reduced density matrix 
$\rho_{s}$ with a target density matrix $\rho_{0}$. In the context of an 
AQO problem, $P_{0}$ is the probability for the quantum annealing (QA) 
processor/chip to be in the GS of the problem Hamiltonian at time $t=T$; 
$\rho_{0} = |E_{g}(T)\rangle\langle E_{g}(T)|$ is the density matrix associated 
with the target state $|E_{g}(T)\rangle$ which is the GS of the problem 
Hamiltonian at $t=T$; and $\rho_{s}$ is the reduced density matrix of the 
QA processor at $t=T$. A study of the performance of adiabatic quantum 
computation using the Uhlmann fidelity as the performance metric has been 
carried out in Ref.~\onlinecite{Dengetal}. We refer the reader to this paper 
for a detailed examination of this question. Still, it is of interest to come to
a better understanding of quantum coherence in the D-Wave hardware 
(viz. phase coherence and entanglement). A demonstration of such coherence
requires a dedicated experimental effort which is currently underway at D-Wave, 
and the results of that work will be reported elsewhere. We stress that the
focus of this paper has been: (i)~a specific application of a QA processor;
and (ii)~the development of techniques which will allow in principle
the solution of arbitrary discrete optimization problems by QA of 
sparsely-connected Ising models.\\